\documentclass[traditabstract]{aa} % for the abstract without structuration 
\usepackage{graphicx}
\usepackage{amsmath}
\usepackage{ulem}
\usepackage{txfonts}
\usepackage{natbib}
\bibpunct{(}{)}{;}{a}{}{,}
\usepackage{float}
\usepackage{calc}
\usepackage{textcomp}
\usepackage{placeins}
\usepackage{color}
\usepackage{rotating} 
\usepackage{lscape}
\usepackage{url}
\usepackage{hyperref}
\usepackage{subfig}
\usepackage[all]{draftcopy}
\usepackage{longtable}
\usepackage{array}
\usepackage{threeparttable}
\usepackage{lscape}
\DeclareTextSymbol{\degre}{OT1}{23}
\newcounter{savedfootnote}

\def \microns{{\,$\mu$m}}
\def \galform{{\sc{galform}}}
\begin{document}
   \title{Constraining the properties of AGN host galaxies with Spectral Energy Distribution modelling.}

\author{L.~Ciesla\inst{1,2},
	V.~Charmandaris\inst{2,1,3},
	A.~Georgakakis\inst{4,2},
	E.~Bernhard\inst{5},
	P. D.~Mitchell\inst{6},
	V.~Buat\inst{7},
	D.~Elbaz\inst{8},
	E.~LeFloc'h\inst{8},
	C. G.~Lacey\inst{6},
	G.~E.~Magdis\inst{9,2},
	and M.~Xilouris\inst{2}.
         }

\institute{	
  University of Crete, Department of Physics, Heraklion 71003, Greece
  \and
  Institute for Astronomy, Astrophysics, Space Applications and Remote Sensing, National Observatory of Athens, GR-15236 Penteli, Greece
  \and
  Chercheur Associ\'e, Observatoire de Paris, F-75014 Paris, France
  \and
  Department of Physics and Astronomy, Max Planck Institut f\"ur Extraterrestrische Physik, Giessenbachstra{\ss}e, D-85748 Garching, Germany
  \and
  Department of Physics \& Astronomy, University of Sheffield, Sheffield S3 7RH, UK		
  \and
  Institute for Computational Cosmology, Department of Physics, University of Durham, South Road, Durham DH1 3LE, UK
  \and
  Aix-Marseille  Universit\'e,  CNRS, LAM (Laboratoire d'Astrophysique de Marseille) UMR7326,  13388, Marseille, France	
  \and
  Laboratoire AIM, CEA/DSM/IRFU, CNRS, Universit\'e Paris-Diderot, 91190 Gif, France	
  \and
  Department of Physics, University of Oxford, Keble Road, Ox- ford OX1 3RH, UK
}	
 
   \date{Received; accepted}

  \abstract
{
Detailed studies of the Spectral Energy Distribution (SED) of normal galaxies are been increasingly used in order to understand the physical mechanism dominating their integrated emission, mainly due to the availability of high quality multiwavelength data from the UV to the far-infrared (FIR).
However, systems hosting dust-enshrouded nuclear starbursts and/or an active galactic nucleus (AGN) due to an accreting supermassive black hole, are especially challenging to study. 
This is due to the complex interplay between the heating by massive stars and the AGN, the absorption and emission of radiation from dust, as well as the presence of the underlying old stellar population.

We use the latest release of CIGALE, a fast state-of-the-art galaxy SED fitting model relying on energy balance, to study in a self consistent manner the influence of an AGN in estimating both the star formation rate (SFR) and stellar mass in galaxies, as well as to calculate the contribution of the AGN to the power output of the host. 
Using the semi analytical galaxy formation model \galform, we create a suite of mock galaxy SEDs using realistic star formation histories (SFH). 
We also add an AGN of Type 1, Type 2, or intermediate type whose contribution to the bolometric luminosity can be variable.

We perform an SED fitting of these catalogues with CIGALE assuming three different SFHs: a single--exponentially--decreasing (1$\tau$-dec), a double--exponentially--decreasing (2$\tau$-dec), and a delayed SFH.

Constraining the overall contribution of an AGN to the total infrared luminosity ($frac_{AGN}$) is very challenging for $frac_{AGN}<$20\%, with uncertainties of $\sim$5--30\% for higher fractions depending on the AGN type, while FIR and sub-mm are essential. 
The AGN power has an impact on the estimation of $M_*$ in Type 1 and intermediate type AGNs but has no effect for galaxies hosting Type 2 AGNs. 
We find that in the absence of AGN emission, the best estimates of $M_*$ are obtained using the 2$\tau$-dec model but at the expense of realistic ages of the stellar population. 
The delayed SFH model provides good estimates of $M_*$ and SFR, with a maximum offset of 10\% as well as better estimates of the age.

Our analysis shows that the underestimation of the SFR increases with $frac_{AGN}$ for Type 1 systems, as well as for low contributions of an intermediate AGN type, but it is quite insensitive to the emission of Type 2 AGNs up to $frac_{AGN}\sim$45\%.

A lack of sampling the FIR, or submm domain yields to a systematic overestimation of the SFR ($<$20\%), independent from the contribution of the AGN. 
Similarly the UV emission is critical in accurately retrieving the $M_*$ for Type 1 and intermediate type AGN, and the SFR of all of the three AGN types.

We show that the presence of AGN emission introduces a scatter to the SFR-$M_*$ main sequence relation derived from SED fitting, which is driven by the uncertainties on $M_*$.

Finally, we use our mock catalogues to test the popular IR SED fitting code \textsc{DecompIR} and show that $frac_{AGN}$ is underestimated but that the SFR is well recovered for Type 1 and intermediate types of AGN. The $frac_{AGN}$, SFR, and $L_{IR}$ estimates of Type 2 AGNs are more problematic due to a FIR emission disagreement between predicted and observed models.

}

  
   \authorrunning{Ciesla et al.}
   \titlerunning{Derivation of stellar mass and SFR in AGN host galaxies}

   \maketitle

%=================================================================================
\section{\label{intro}Introduction}
	
The formation  of galaxies  and their evolution  with cosmic  time are open  problems in current  astrophysical research.   
Understanding the assembly of galaxies and the  built-up of their stellar populations is challenging  because  the  relevant  physical processes  are  complex, interconnected and  operate on a  large range of scales. 
Gas inflows from  the intergalactic medium  (IGM) for  example, are  important for supplying galaxies with  fresh material that can be  turned into stars or feed  supermassive black holes  (SMBH) at their  centers. Feedback processes associated with stellar evolution of Active Galactic Nuclei  (AGN) can drastically  modify the  physical conditions  of the Inter-Stellar  Medium (ISM)  thereby, affecting  the formation  of new stars.   
Supernovae explosions  enrich with  heavy metals  the gaseous component of galaxies and  their environments and therefore change the composition  of  the ISM  with  implications  on star-formation.   
The density of  IGM on large  scales or interactions with  nearby galaxies also have  an impact on the  ISM of individual systems  and can modify their evolutionary path.

One approach for  shedding light into the physics  of galaxy formation and  evolution is  population studies.   
Multi-wavelength observations provide  information  on  galaxy  properties  such  as  stellar  mass, star-formation history  (SFH), gas content,  AGN activity, kinematics, structural parameters  or position on  the cosmic web. 
Each  of these observationally   determined  parameters   probe   different  physical processes.  
Exploring correlations between  them for large samples can therefore provides  insights on  how galaxies form  and evolve  in the Universe.  
For example, the discovery of the bimodality of galaxies on the colour (proxy to star-formation history) versus stellar-mass plane \citep[e.g.][]{Strateva01,Baldry04,Bell04,Baldry06} has been interpreted  as evidence for  star-formation  quenching  that may  be driven    by   either   internal    (e.g.   feedback) or external (e.g. environment) processes \citep[e.g.][]{Bell04,Faber07,Schawinski14}.   
The star-formation main sequence (MS),  i.e. the relatively tight  correlation between stellar mass   and  star-formation   rate  (SFR)   of   star-forming  galaxies \citep[e.g.][]{Salim07,Noeske07_SFseq,Elbaz07,Rodighiero11,Speagle14}, suggests  that the bulk  of the stars in  the Universe  form via  secular processes  rather than  in violent events,  such  as  mergers.    
Addtionally,  the  slope  and  redshift evolution of  the MS  normalization has been  discussed in  context of feedback processes  and gas exhaustion with  cosmic time, respectively \citep[e.g.][]{Noeske07_gas,Zheng07,Tacconi13,  Guo13}.   
Correlations between   galaxy   structural  parameters,   SFR   and  stellar   mass \cite[e.g.][]{Kauffmann04,Wuyts11,Bell12,Lang14}   suggest   that  the formation  of galaxy bulges  and the  quenching of  star formation are likely related to the same underlying processes.

Among  the   different  galaxy  properties  that   are  accessible  to observations,  the stellar  mass and  SFR  play an  important role  in galaxy evolution  studies.  
This  is not surprising.   
Both quantities provide a measure of the SFH of the galaxy, either integrated over its lifetime  (stellar mass)  or  averaged in  the  last few  tens to  few hundred  million  years   (instantaneous  SFR).   
Indeed,  all  galaxy properties show  strong correlations with either stellar  mass, SFR or both  \citep[e.g.,][]{Kauffmann04}.   
This has  led  to an  increasing refinement in  methodology to provide more accurate  and less biased observational constraints  to the stellar  mass and SFR  of individual galaxies.   
Among the  different  approaches, the  one  that has  been extensively  used  in the  literature  is  the  use of stellar  population synthesis    models   \citep[e.g.][]{BruzualCharlot03,Maraston05}   to generate   Spectral Energy  Distribution (SED) templates  for different star-formation  histories   and  then  fit  them   to  the  broad-band photometry      of      galaxies      \citep[e.g.][and      references therein]{Walcher11}.  
This is driven  by the explosion in recent years in the quality and quantity of multi-wavelength imaging surveys, which provide UV to  infrared (IR) SEDs of large galaxy  samples over a wide range of redshifts.  
This approach for inferring galaxy parameters has also  been extensively  tested to  identify limitations  and potential sources   of   systematics  \citep[e.g.][]{Pozzetti07,   Marchesini09, Conroy09,   Wuyts09,   Ilbert10,Michalowski12,   Pforr12,   Banerji13, Schaerer13, Mitchell13, Buat14}.

The  overall  consensus  is   that  stellar  masses  can  be  reliably constrained,  although systematics  up to  the 0.5\,dex  level remain, depending  on the  adopted Initial  Mass Function,  stellar population libraries, functional form of the  model SFH and the implementation of dust attenuation.   
The SFR of  galaxies is often determined  using as tracers    the   observed    luminosity   at    specific   wavelengths \cite[e.g.][]{Kennicutt98,Hopkins04}.  
Nevertheless,  studies that fit the full SED  to derive SFR find an  overall reasonable agreement with estimates          based          on         specific          tracers \citep[e.g.][]{Salim07,Salim09,Walcher08}.

One  aspect of  galaxy evolution  that has  developed  considerably in recent years is the relation  to SMBH growth.  
This has been motivated by the tight correlations between proxies of the stellar mass of local spheroids  and   the  mass  of   the  black  hole  at   their  centres \citep[e.g.][and references  therein]{Kormendy13}.  
One interpretation of these correlations is that  of the co-evolution of AGN and galaxies dictated by a common gas reservoir that forms new stars and also feeds the central  black hole.   
An alternative explanation  is that  of AGN outflows that  heat and/or  expel the cold  gas component  of galaxies thereby,  regulating the  formation of  new stars  and  ultimately the accretion        onto       the       central        SMBH       itself \citep[e.g][]{Silk98,Fabian99,King03,King05,DiMatteo05}.    
There   is indeed increasing observational evidence  for AGN-driven winds at both the        local        Universe        and       high        redshift \citep[e.g][]{Crenshaw03,Blustin05,Tombesi10,Tombesi12, SaezChartas11,Lanzuisi12,Cicone12,Cicone14,Harrison14}.   
What remains controversial however,  is how common  these outflows are and  if they are energetically important to affect the ISM of their hosts.  
One way to  approach these  issues  and place  AGNs  in the  context of  galaxy evolution  is statistical  studies of  the host  galaxy  properties of large  AGN  samples.   
Questions  that  this  approach  could  address include,  when during  the lifetime  of  a galaxy  accretion onto  the central  SMBH  is triggered,  how  black  hole  growth is  related  to star-formation, and whether AGNs affect their host galaxy properties.

Multi-wavelength  survey  programs in  the  last  decade have  started addressing  these questions.   
Far-infrared (FIR)  and  sub-mm surveys with {\it Herschel} for example, measure the mean SFR of AGN hosts via stacking  methods and  show  that AGN,  on  average, lie  on the  main sequence star-formation        at all redshifts \citep[e.g.][]{Mullaney12,Santini12,Rosario12,Rosario13}.  
 This finding suggests that at least  in an average sense the  same physical processes govern the formation  of stars in galaxies  and the growth of  black holes at their  centers  \citep{Silverman09,Georgakakis11,Mullaney12}.  
 Studies of the star-formation properties  of individual AGN hosts (rather than averages  over  populations)  suggest   a  wide  range  of  SFH.   
 The rest-frame           colors           of           AGN           hosts \citep[][]{Aird12,Georgakakis14,Azadi14}  or  their  position  on  the color--magnitude diagram \citep[e.g.][]{Sanchez04,Nandra07,Schawinski09} indicate that they are scattered at  the red sequence  of passive galaxies,  the star-forming cloud  and the green  valley in  between.  
 A  higher incidence  of AGN among   green   valley   galaxies  is   claimed   \citep[][]{Nandra07,Schawinski09},  i.e.  systems  with  colors in  the transition  region between the red sequence and  the blue cloud.  This can be interpreted as  evidence   for  AGN  being   responsible  for  the   quenching  of star-formation  in  galaxies and  hence,  their  transition from  blue star-forming to red  and dead systems.  
 At the  same time however, the importance   of   constructing   appropriate  control   samples   when determining  the fraction of  AGN among  galaxies is  also emphasized.
The evidence for an increased AGN fraction among green valley galaxies is less  strong, or may  even disappear once  the stellar mass  of AGN hosts    and    the     control    galaxy    samples    are    matched \citep[e.g.][]{Silverman09,Xue10,Aird12}.  
Constraints  on the stellar ages  of  AGN  hosts   further  suggests  that  black  hole  accretion preferentially occurs few hundred  to few thousand million years after the peak  of  star-formation \citep[][]{Kauffmann03,Wild10,HernanCaballero14}, i.e. during a period when    galaxies     can    be    identified     as    post-starbursts \citep[e.g.][]{Goto06,Georgakakis08}. 
A time  lag between the peak of star-formation  and  the black  hole  growth  may  pose a  problem  to AGN-driven star-formation quenching scenarios.

The  studies  above  demonstrate   the  importance  of  measuring  the star-formation and stellar  mass for AGN hosts.  
This  is not only for understanding  the relation between  star-formation history  and black hole  growth   in  individual  systems,  but   also  for  constructing appropriate  control  samples  of  inactive (non-AGN)  galaxies.  
The determination of stellar  masses or SFR for AGN  hosts is challenging.
Emission from the central engine contaminates or may even dominate the underlying galaxy light, thereby rendering the determination of galaxy properties,  via  e.g.  SED  fitting  methods,  difficult.   
Different approaches have been adopted in  the literature to address this issue.
Standard  SED fitting  methods  are often  applied  to determine  host galaxy properties only for low  luminosity and obscured AGN, under the reasonable assumption that contamination in these objects is small.  
A potential problem with this approach  is that under certain models for the   co-evolution  of   AGN  and   galaxies  \citep[e.g][]{Hopkins06} different levels  of obscuration and  different accretion luminosities correspond to different evolutionary  phases of the black hole growth. 
The  selection against  unobscured  and luminous  AGN might  introduce biases and lead to erroneous conclusions on the relation between black hole accretion  and galaxy formation.   
An alternative approach  is to add  AGN templates  to the  stellar template  library and  perform SED deconvolution   to   separate    the   galaxy   and   AGN   components \citep[e.g.][]{Bongiorno12,Lusso11,Lusso13,  Rovilos14}.  
This  can be powerful,  although degenerations between  AGN and  galaxy  templates  may introduce systematics in the determination of host galaxy parameters.

To  date there are  still few  studies that  explore and  quantify the impact  of AGN contamination  on the  determination of  the underlying host galaxy properties via SED fitting \citep{Wuyts11,HaywardSmith14}.
In this  paper, we address  this using the bayesian-based  SED fitting code CIGALE  \citep[][Burgarella et al.,  in prep; Boquien et  al., in prep]{Noll09}.  
To   do  so,  we  follow  the   method  developed  in \cite{Mitchell13}   and   use   simulated   galaxies  from   the   SAM (Semi-Analytica Model) code \galform\ from which  we know the exact value of  the stellar mass and SFR.  
After  building the UV-to-submm  SEDs of the  \galform\ objects, and adding an AGN contribution,  we will study our ability to retrieve the  original  properties using  CIGALE.

The paper is organized as  follows.  First, we introduce \galform\ and present the derived star formation history (SFH) of a hundred galaxies at  $z\,=\,1$ (Section~\ref{build}).   
We  describe how  we build  our mocks   SEDs    using   the    modeling   function   of    CIGALE   in Section~\ref{build},   and  how   we  perform   the  SED   fitting  in Section~\ref{sedfit_}.   
In  the same  section,  we  also present  the comparison between the true values  of the stellar masses and SFRs and the  outputs of  CIGALE  in normal  galaxies  as well  as  in AGN  host galaxies.  
Finally, we  discuss  the  impact of  our  results on  the SFR--$M_*$  relation in Section~\ref{ms}.   
In Section~\ref{DecompIR}, we use our mock samples to  evaluate the performance of the popular IR SED fitting code \textsc{DecompIR} \citep{Mullaney11}.  
The purpose of this  work is  not to  test thoroughly  the validity  of  AGN emission models,  neither  the  ability  of  SED fitting  codes  to  accurately retrieve  the properties of  the AGN  through its  IR emission  but to
analyze the  impact of  this emission on  the derivation of  the basic host galaxy properties.
	 
In this work,  we assume that $\Omega_m =$  0.25, $\Omega_{\Lambda} =$ 0.75, and $H_0\,=$\,73\,km s$^{-1}$ Mpc$^{-1}$.  
These values are used because the \galform\ model is  built on top of merger trees generated from  the  original  Millenium  simulation  \citep{Springel05},  which
assumes  a WMAP1  like cosmology.   
This  choice does  not impact  our results, as we only use  this cosmology to convert physical quantities to  observables, without comparisons  with other  works based  on more recent cosmological  parameters.  All of  the stellar masses  and SFRs are provided assuming a IMF of Salpeter.

%=================================================================================
\section{\label{build}Building a realistic mock galaxy sample}

In this paper  we explore the problem of AGN/stellar light decomposition  via SED  template fits  to  multi-wavelength broad-band photometric  data.   
The goal  is  to  quantify  how well  the  galaxy properties of AGN hosts, such as stellar mass and star-formation rate, can be  measured from  broad-band photometry. 
Our  approach is  to use mock  galaxies extracted  from the  \galform\ SAM  \citep{Cole00}, for which  the SFHs,  stellar masses  ($M_*$) and  instantaneous  SFR are known.  
This information  allows the construction of UV  to submm SEDs for the  mock galaxies.   
AGN templates are  then added to  the galaxy emission to generate composite  SEDs. 
These are then integrated within broad-band  filters  to generated  mock  photometric catalogues.   
The fitting modules of the CIGALE code are then applied to the mock galaxy photometry to decompose  AGN from stellar light. 
Each  of these steps are described in the following sections.

%=================================================================================
	\subsection{\label{sfh}Simulations of realistic star formation histories}

	To  generate our  grid  of  mock SFHs,  we  use the  \galform\ semi-analytic galaxy  formation model to simulate the  assembly of the galaxy  population within  the context  of the  $\Lambda$CDM  model of
	structure  formation  \citep{Cole00,Bower06,Benson10}.   
	The model  is constructed on top of dark matter halo merger trees extracted from the Millennium  dark matter  N-body simulation  \citep{Springel05}. 
	Within each halo,  the baryonic mass is  divided into hot and  cold gas along with stellar disk  and bulge components.  
	The model  then solves a set of coupled differential equations  that describe how mass is exchanged between these  different components.  
	Star  formation in the  model is split into  quiescent star formation  that occurs in galaxy  disks and bursts of star formation that are triggered by galaxy mergers and disk instabilities \citep{Baugh05,Bower06,Lagos11}.

	Star formation  histories are extracted  from the fiducial  version of the model presented by  \cite{Mitchell14}.  
	We randomly select a total of 100  galaxies from that  simulation at redshift $z=1$.
	Motivated  by  observational evidence  that  AGN  hosts are  typically massive and  lie, at least in  average sense, on the  main sequence of star-formation \citep[e.g.,][]{Mullaney12,Santini12,Rosario12,Rosario13}, we 	also choose the mock galaxies to sit on the main sequence of the  SAM  at redshift  $z=1$  and to  have  high  stellar masses.   
	In particular they fulfill the following criteria:

	\begin{enumerate}
		\item    Specific   Star-formation    rates   sSFR$>$ 0.1\,Gyr$^{-1}$. This cut  separates main star-formation sequence from passive galaxies in the model at $z=1$ \citep{Mitchell14}.
		\item    Stellar   mass   in  the   range   $10<   \log (M_*/\rm{M_{\odot}}) <11$.
	\end{enumerate}

\begin{figure*}%[!h] 
  		\includegraphics[width=\textwidth]{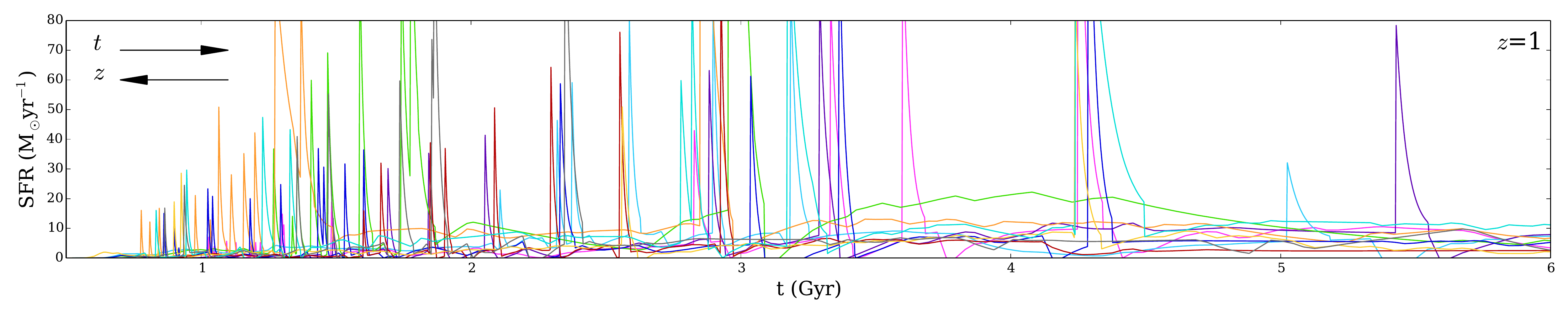}
  		\caption{ \label{sfh_plot} Examples of 10 SFHs of simulated galaxies extracted from the \galform\ SAM.}
	\end{figure*}

\begin{figure}%[!h] 
 		 \includegraphics[width=\columnwidth]{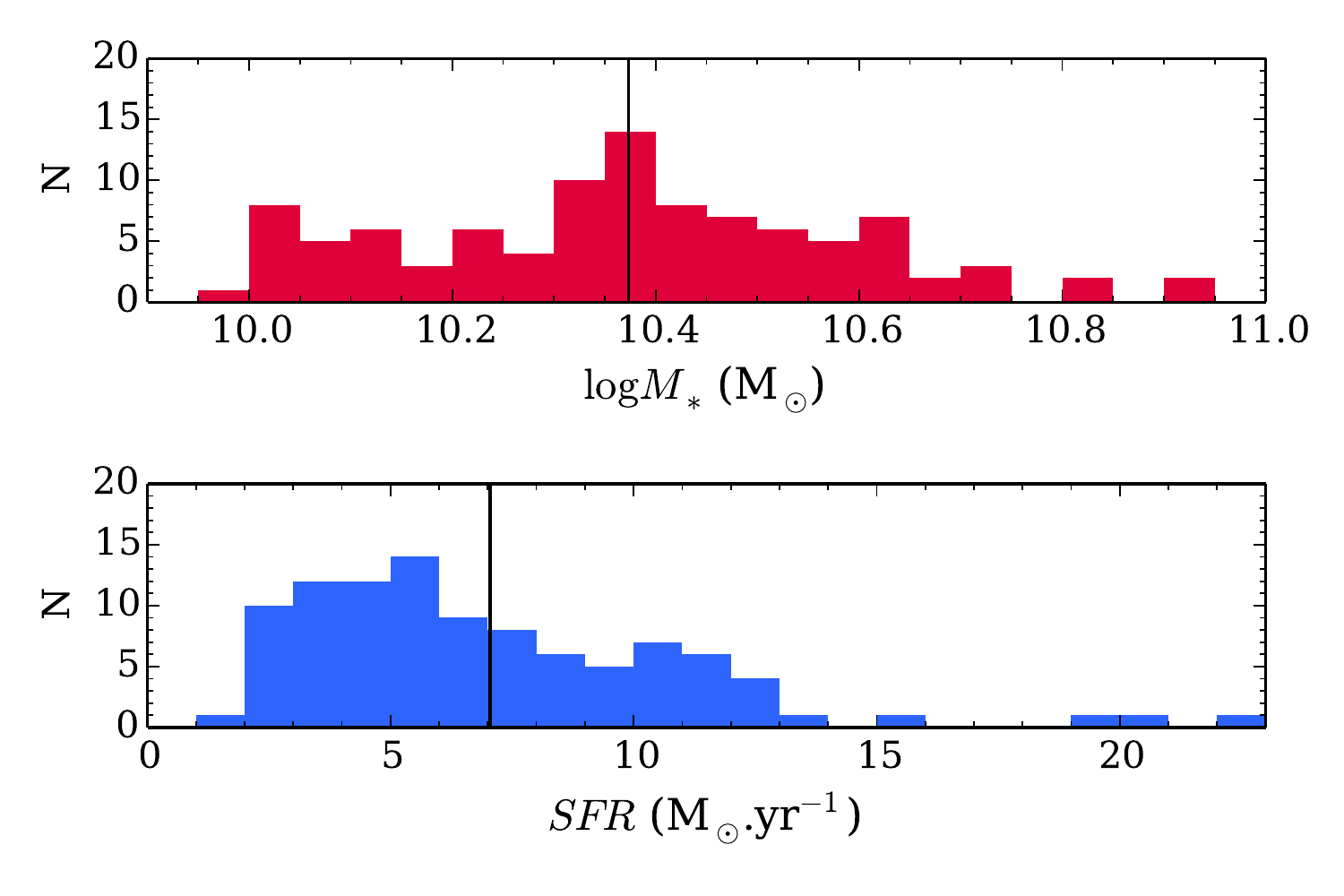}
  		\caption{  \label{histmstarsfr}   Distribution  of  $M_*$   and  SFR associated  with the  \galform\ SFHs.  The black  lines show  the mean value of the distributions.}
	\end{figure}

	\noindent For  these star  formation histories, we  sum over  the star formation in all  progenitors of the final galaxy  and we also combine the stellar mass  assembly of the disk and  bulge together.  
	Bursts of star formation in the model can occur over relatively short timescales in some  cases and so we  construct star formation  histories from the model to have high temporal resolution.  
	We show ten examples of the produced SFHs  in Figure~\ref{sfh_plot}.  
	The  distribution of stellar mass  and SFR at  redshft $z=1$ of the 100 simulated galaxies selected  for our analysis are presented in Figure~\ref{histmstarsfr}.   
	The mean  values of the sample are $\log M_*$=10.37\,M$_{\odot}$ and SFR=7.05\,$M_*$.yr$^{-1}$.
		
	%\FloatBarrier
	
	\subsection{\label{sedmod}Simulations of UV to submm SEDs}
	
	\begin{figure}%[!h] 
  		\includegraphics[width=\columnwidth]{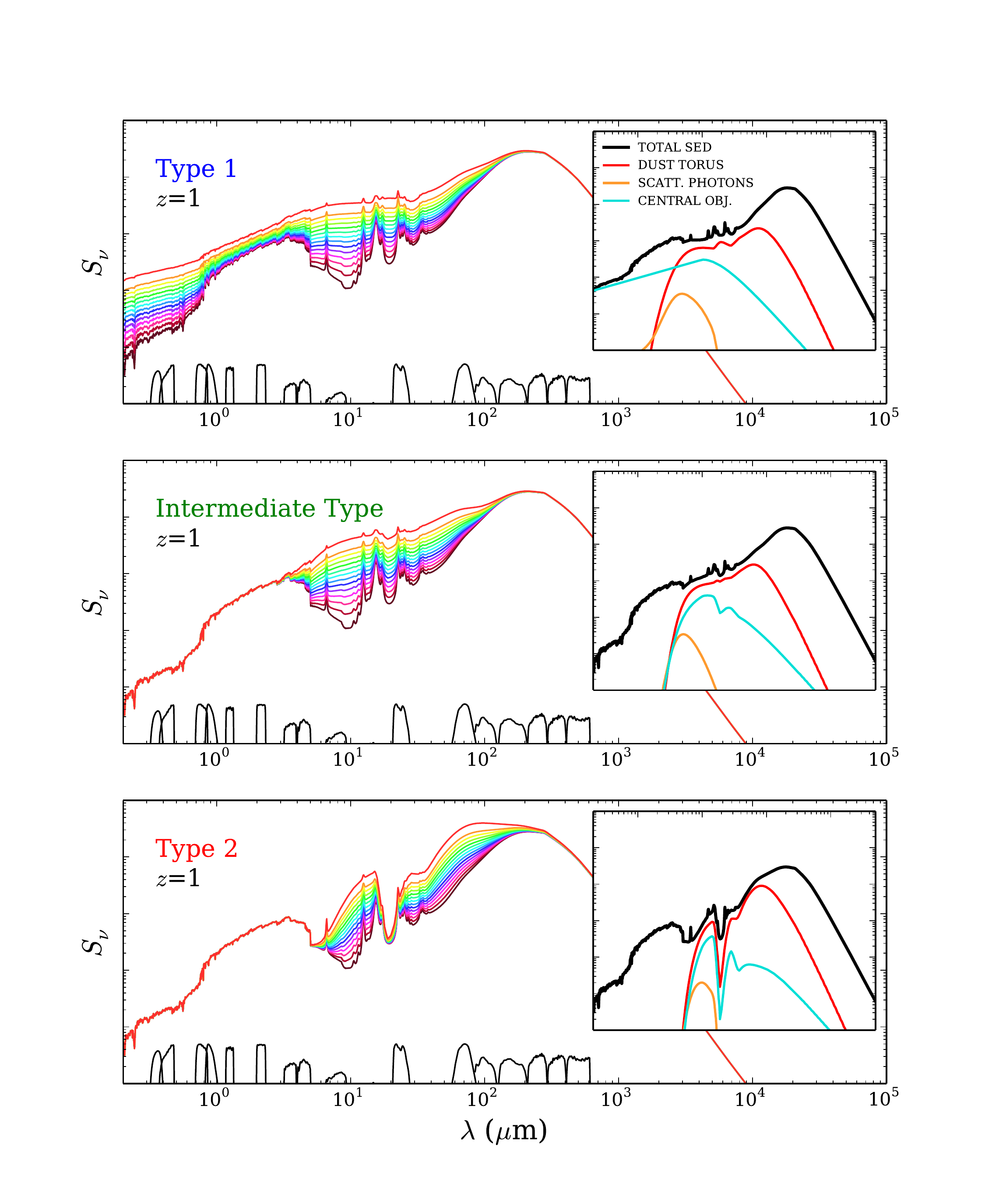}
  		\caption{ \label{inputsed} SEDs generated by GIGALE using a particular SFH extracted from \galform\ and adding on that different AGN templates with different normalizations relative to the galaxy light. 		Each panel corresponds to one of the three Fritz et al. (2006) templates presented in Table \ref{mockparam}: the upper panel is for Type 1 AGN, the middle panel is for the  Intermediate AGN type, and 		the lower 	panel is for Type 2 AGN. SEDs are color-coded according to the contribution of the AGN to the total IR luminosity. The black solid lines at the bottom of each panel are the broad-band filters of Table~\ref{filt}, within which the model SEDs are  integrated to generated mock photometric catalogues. The inset plots show the contribution of the three AGN components to the total SED for $frac_{AGN}$=40\%. Red is for the dust torus emission, orange is the scattering component, and cyan the direct emission from the central AGN.}
	\end{figure}	

	\begin{figure}%[!h] 
 		 \includegraphics[width=\columnwidth]{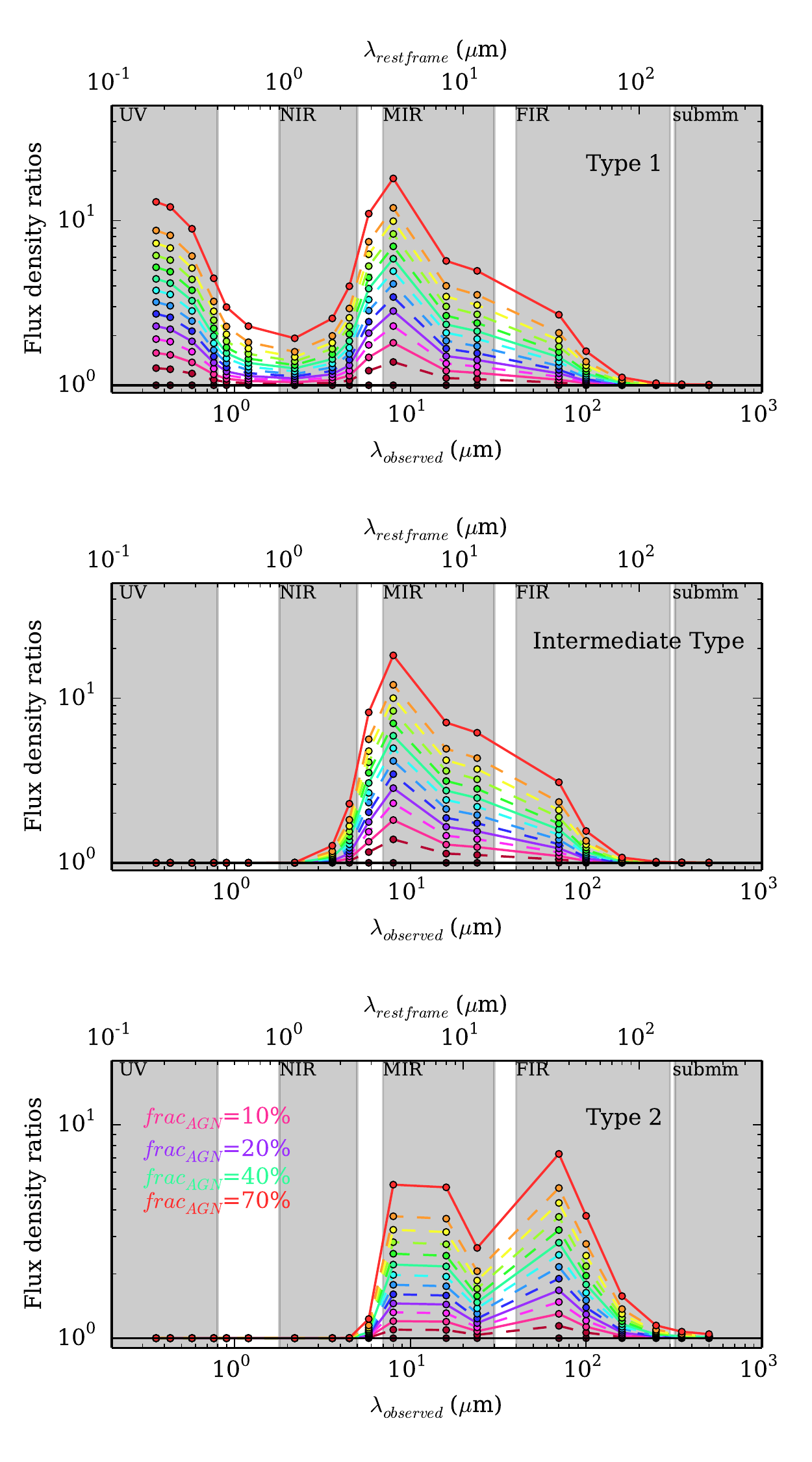}
 		 \caption{ \label{fluxratios} The SEDs of Figure~\ref{inputsed} normalized to the SED with no AGN emission, i.e. $frac_{AGN}=0$. The three panels correspond to one of the Fritz et al. (2006) templates presented in Table \ref{mockparam}: upper panel is for Type 1 AGN, the middle panel corresponds to the intermediate AGN type, and the lower panel is for the Type 2 AGN. The grey regions indicated different rest frame spectral domains from UV to sub-mm. Normalized SEDs are color-coded according to the $frac_{AGN}$ parameter. Solid colored lines indicate specific SEDs, i.e. where $frac_{AGN}$=10\% (pink), 20\% (purple), 40\% (green), and 70\% (red). }
		\end{figure}

	CIGALE\footnote{The code is freely available at: \url{http://cigale.lam.fr/}}  is  a   package  that  has  two  different  and independent functions:  an SED modeling  function (Boquien et  al., in prep) and an  SED fitting function (Burgarella et  al., in prep).  
	The baseline functions of CIGALE are  presented  in  \cite{Noll09}.   
	Based on the same general principles as the original version of CIGALE, this entirely new version has been designed for a broader set of scientific applications and better performance.	
	The latest  version of the  modeling  function  of   CIGALE  is  briefly  described  in  this section.

	The  SED   modeling  function  of   CIGALE allows  the construction of  galaxy SEDs from  the UV to  the submm by  assuming a stellar  population library and  star-formation histories  provided by the  user.  	CIGALE builds  the  SED taking  into account  energy balance, i.e., the energy absorbed  by dust in UV-optical is reemitted in the IR.
	
	In CIGALE, the SFH can be  handled in two different ways. 
	The  first, is  to  model  it using  simple  analytic functions  (e.g. exponential forms, delayed  SFHs, etc).  
	The second is  to provide more complex (non-analytic) SFHs \citep[e.g.,][]{Boquien14}, such as  those provided by  the \galform\ SAM.   
	The stellar  population models  of either  \cite{Maraston05} or \cite{BruzualCharlot03} are convolved with  the adopted SFH to produce stellar SEDs, which are then  attenuated by dust.  
	The energy absorbed by the  dust is reemitted in the  IR using a choice  of different dust templates \citep{DaleHelou02, Dale14, DraineLi07, Casey12}.

	CIGALE also allows the  emission from AGN to  be added to the  stellar SED.   
	The  AGN templates from the library of \cite{Fritz06} are adopted.  
	These SEDs consist of two  components.  
	The  first  one  is the  isotropic  emission of  the central source, which is assumed to be point-like.  
	This emission is a composition  of power  laws with  variable indices  in  the wavelength range  of 0.001-20\microns.   
	The  second component  of  the \cite{Fritz06} models  is radiation from dust with  a toroidal geometry in the vicinity  of the central engine.   
	Part of the  direct emission of the AGN  is either absorbed by  the toroidal obscurer  and re-emitted at longer  wavelength (1-1000\microns)  or scattered  by the  same medium.
	 Dust  can  be optically  thick to its  own radiation,  thus requiring  the numerical resolution  of  the  radiative  transfer problem.   
	 In  \cite{Fritz06} models, the conservation  of energy is always verified  within 1\% for typical solutions,  and up to  10\% in the  case of very  high optical depth and  non-constant dust density.  
	 The choice of adding  the \cite{Fritz06} library into CIGALE  is driven by the energy balance handling of  the two components, which also matches the  energy  conservation  philosophy  of CIGALE.
	 Furthermore,  this library  has  been  tested  in  numerous  studies  of  the  literature \citep[e.g.,][]{Fritz06,      Hatziminaoglou08,      Hatziminaoglou10, Feltre12}.
		
	The relative normalization of these  AGN components to the host galaxy SED is handled through a parameter  which is the fraction of the total IR luminosity due to the AGN so that

	\begin{equation}
    		L_{IR}^{AGN} = frac_{AGN} \times L_{IR}^{TOT},
  	\end{equation}
	\noindent where $L_{IR}^{AGN}$ is  the AGN IR luminosity, $frac_{AGN}$ is  the   contribution  of  the   AGN  to  the  total   IR  luminosity ($L_{IR}^{TOT}$), i.e. $L_{IR}^{starburst}+L_{IR}^{AGN}$.
	Thus, estimating $L_{IR}^{AGN}$ depends on the constraints on  $frac_{AGN}$.

	 For  each SFH  provided by  \galform, we  compute  a host galaxy SED  using the \cite{Maraston05} stellar  population models and assuming  the   \cite{Calzetti00}  extinction  law.    
	 The  amount  of reddening, $E(B-V)_*$,  is chosen to be 0.2,  motived by observational studies  of AGN host  galaxies \citep{Kauffmann03a,  Hainline12}.  
	 The energy absorbed in UV-Opt-NIR is re-injected in IR, providing the normalization of the \cite{DraineLi07} models in order to maintain the energy balance of the SED.    
	 The   template  is   chosen   following   the  results   of \cite{Magdis12} who  fit high-$z$ galaxies  with the \cite{DraineLi07} models.  
	 The  parameters used to  model the galaxies are  presented in Table~\ref{mockparam}.

	We   select  three  AGN   model  templates   from  the \cite{Fritz06} library to be added  to the galaxy SEDs.  
	These include a Type 1  AGN (i.e.  unobscured), a Type 2 AGN  (i.e.  obscured) and a template  that lies  in-between the  first two  and is  referred  to as intermediate  type.  
	The  latter model  displays a  power-law spectral shape  in the  mid-IR  without strong  UV/optical emission.   
	Previous studies  indicate  that  such  an intermediate  template  is  indeed necessary  to represent  the diversity  of  the observed  SEDs of  AGN \citep[][]{Hatziminaoglou08,Hatziminaoglou09,Feltre12}.    		The \cite{Fritz06} model parameters for  the three templates are presented in Table~\ref{mockparam}\footnote{We  note an error  in the definition of the angle  relative to the line of  sight $\psi$ in \cite{Fritz06}: 	$\psi$=0$^{\circ}$  corresponds  to a  Type  2  AGN  whereas an  angle $\psi$=90$^{\circ}$ is for a Type  1 (J. Fritz, private communication).}.  
	The parameters for the Type 1 and Type 2  AGN templates are representative of  local unobscured and obscured AGN, respectively  \citep{Fritz06}.  
	For the intermediate AGN type we  adopt one  of the  \cite{Fritz06}   models with  low equatorial optical  depth   \citep[][Buat  et  al.,   in  prep]{Hatziminaoglou08,Hatziminaoglou09,Feltre12}.  
	The three  AGN templates adopted in this paper  are  the minimum  required  to  represent  the variety  of  the broad-band  AGN SEDs.  
	There are  strong degeneracies  among different parameters  of the  \cite{Fritz06}  library  that cannot  be broken  by multiwavelength   photometric   data   alone.   
	Additional   parameters combinations   to  those   shown  in   Table~\ref{mockparam}   do  not necessarily  result  in AGN  broad-band  SEDs  that are  distinctively different from  our Type  1, Type 2  and 		intermediate type  models. 
	In Figure~\ref{inputsed}, we  show the SEDs  corresponding to  one of the SFHs presented in Figure~\ref{sfh_plot}.

	The  modeled SEDs  are integrated  within the  broad band filters  of Table~\ref{filt} to  produce mock  photometric catalogues.
	These will  be used in  the next section  to assess the  reliability of AGN/galaxy decomposition using photometric data and quantify the level of accuracy at which physical parameters  of the underlying galaxy 	can be derived.
	Random noise  is added  to the  fluxes  of each source assuming  Gaussian errors with  standard deviation 10\%  of the flux  in a given  waveband.  
	We associate a photometric error of 15\% with all the flux densities.
	The broad-band  filters  of  Table~\ref{filt}   sample  a  wide  range  of wavelengths, as shown on  Figure~\ref{inputsed}.  
	In order to quantify the  level of  AGN  contribution in  each  of these  filters,  we show  in Figure~\ref{fluxratios}  the  flux density  ratios  of  the SEDs  that include   an   AGN  component   relative   to   those   that  	do   not ($frac_{AGN}=0$). 
	A Type  1 AGN with a $frac_{AGN}=10\%$  will have an AGN emission higher by a factor of 2 in the UV and MIR  rest frame compared to the same   SED   without   an   AGN   component.   
	In   UV  rest   frame,   a   fraction $frac_{AGN}=40\%$  is  sufficient to dominate the  emission, with an AGN  contribution four times higher than the young  stellar population emission.  
	In MIR  rest   frame, the AGN emission  is dominant for $frac_{AGN}>$20\% with an emission three times higher than the stellar emission.   
	The UV and the  MIR are thus key  domains to perform AGN/galaxy decomposition in the case of a Type 1 AGN SED as also shown in previous work \citep[e.g.,][]{Weedman04,Wu09}.  
	The AGN emission of the intermediate type is  visible in the MIR domains, especially at 4\microns\  rest frame, where it is brighter  than the host galaxy by a factor  of 2 for a fraction of 10\%.   
	For the Type  2 AGN,  it  is clear  that the  emission  of the  AGN cannot  be detected  below  2\microns\  rest   frame. 
	In this  model, the torus is optically  thick and the emission from  the  inner,  hotter  part  of the  torus,  emitting  at  shorter wavelength is  completely absorbed.  
	In FIR rest frame,  the  emission of the AGN contributing to 70\% of the total $L_{IR}$ will dominate the emission of the host galaxy by a factor  of 7. 
	Given the ratio between the AGN emission and the host galaxy emission (Figure~\ref{fluxratios}), two bands seem to  be the key  to constrain the  Type 2 AGN  emission, the 3-10\microns\ rest frame and  the 	30-40\microns\ rest frame as already noticed in previous works \citep[e.g.,][]{Laurent00}.  
	However, we note that dust emission templates are not well constrained in the 30-40\microns\ range \citep{Ciesla14} and  thus improving them can help disentangling the AGN contribution in the FIR.

	\begin{table*}
		\centering
		\caption{Galaxy and AGN parameters adopted to generate
                  mock galaxy SEDs.}
		\begin{tabular}{l c c}
	 	\hline\hline
		Parameter & Value  & Description\\ 
		\hline
		\multicolumn{3}{c}{Dust attenuation}\\  
		\hline
		\\
		$E(B-V)_*$ 		&  0.2	\\
		\hline		
		\multicolumn{3}{c}{Dust template: \cite{DraineLi07}}\\  
		\hline
		\\
		$q_{PAH}$	(\%)	& 	3.19 & Mass fraction of PAH to the total dust mass. \\
		$U_{min}$		& 	8.0 & Min. intensity of the interstellar radiation field.\\
		$U_{max}$		& 	$10^6$& Max. intensity of the interstellar radiation field.\\
		$\gamma$	(\%)	& 	2 & Relative contribution between dust heated in photodissociation\\
		 				& 	  &  regions, and dust heated by diffuse stellar population.\\
		\hline		
		\multicolumn{3}{c}{AGN emission}\\  
		\hline
		\\
		$R_{max}/R_{min}$ &  60 & Ratio between outer and inner radius of the torus.\\
		$\tau_{9.7}$ & 1.0 (for int. type) & Optical depth at 9.7\microns.\\	
				 & 6.0 (for Type 1 \& Type 2)\\		
		$\beta$ & -0.5 & Linked to the radial dust distribution in the torus.\\		
		$\gamma$ &  0.0 & Linked to the angular dust distribution in the torus.\\
		$\psi$ & 0.001 (for int. type \& Type 2) & Angle with line of sight.\\
		 		& 89.9 (for Type 1)\\
		$\theta$ & 100 & Angular opening angle of the torus.\\
		$frac_{AGN}$& 0., 0.05, 0.1, 0.15, 0.2, 0.25, 0.3, & Contribution of the AGN to the total $L_{IR}$.\\
					     & 0.35, 0.4, 0.45, 0.5, 0.55, 0.6, 0.7 \\ 
		\hline
		\label{mockparam}
		\end{tabular}
	\end{table*}

	\begin{table}
		\centering
		\caption{Broad-band filter-set used in this
                  paper. }
		\begin{tabular}{ l l c }
	 	\hline\hline
		Telescope/Camera & Filter Name & $\lambda_{mean}$(\microns) \\ 
		\hline
		MOSAIC & U & 0.358 \\
		HST		& ACS435 & 0.431 \\
			& ACS606 & 0.573 \\
			& ACS775 & 0.762 \\
			& ACS850 & 0.9 \\
		Subaru/MOIRCS &  J & 1.2 \\
		CFHT/WIRCam& Ks & 2.2 \\
		\textit{Spitzer}& IRAC1& 3.6 \\
			& IRAC2 & 4.5 \\
			& IRAC3 & 5.8 \\
			& IRAC4 & 8 \\
			& IRS16 & 16 \\
			& MIPS1 & 24 \\
			& MIPS2 & 70 \\
		\textit{Herschel}& PACS green & 100 \\
			& PACS red & 160 \\
			& PSW & 250 \\
			& PMW & 350 \\
			& PLW & 500 \\
		\hline
		\label{filt}
		\end{tabular}
	\end{table}

%=================================================================================	
\section{\label{sedfit_}Recovering the mock galaxy properties}

The SED  fitting functions of  CIGALE are applied  to the mock photometric galaxy catalogue  of the previous section to separate the  stellar  emission from  the  AGN  component  and investigate  how accurately stellar  masses and star-formation rates  can be determined for galaxies that host an AGN.  

To perform  the SED fitting  analysis, CIGALE first builds  models corresponding to a range of input parameters for both the stellar  and AGN components.   
The adopted parameters used  in the fitting  procedure are presented  in Table~\ref{fitparam}.   
The ones related to  the galaxy host emission templates are selected based  on the experience gained  from galaxy SED modeling at  intermediate and high redshift using CIGALE \citep[e.g.,][]{Giovannoli11,    Buat12, Burgarella13, Buat14}.

For the  building of a galaxy template  SED, it is  first necessary to make some assumptions  about the  star-formation  history,  which will  be convolved  with  the  stellar  libraries  to yield  galaxy  SEDs.   
The SFH  of real galaxies  are expected to  be highly stochastic.  
It  is therefore impractical and probably meaningless to  assume complex  SFHs like those  shown in Figure~\ref{sfh_plot}, when fitting  multiwavelength  photometric  data.   
It is  common  practice instead,  to assume  simple functional  forms, such  as an exponentially  decreasing  SFR \citep[1$\tau$-dec,   e.g.,][]{Ilbert13,Muzzin13},   an  exponentially increasing   SFR   \citep[1-exp-ris,  e.g.,][]{Pforr12,Reddy12},   two exponential  decreasing  SFR   laws  with  different  e-folding  times \citep[2$\tau$-dec,  e.g.,][]{Papovich01,Borch06,Gawiser07,Lee09}, a delayed  SFR \citep[e.g.,][]{Lee10,Lee11,Schaerer13}, or a lognormal SFH \citep{Gladders13}.  
We  consider in this work  the 1$\tau$-dec, the  2$\tau$-dec, and the  delayed models.
First tests made with CIGALE on the lognormal SFH show that the parameters associated are not constrained by broad band photometry.
However, as demonstrated in Figure~2 of \cite{Gladders13}, only specific combinations of the parameters of this SFH lead to a non null instantaneous SFR at the age of the galaxy.
Thus, we do not consider here the lognormal SFH as it provides problematic SFRs.
Furthermore, since the 1-exp-ris model is  recommended to only model the SFH of galaxies at $z>$2 \citep[e.g.,][]{Maraston10,Papovich11,Pforr12,Reddy12},  we do not test it  in this work. 

The 1$\tau$-dec is represented by the following equation:

\begin{equation}
SFR(t) \propto exp(-t/\tau_1) 
\end{equation}

\noindent where $t$ is the time and $\tau_{1}$ the e-folding time of the old stellar population.
The 2$\tau$-dec is obtained adding a late burst to the 1$\tau$-dec SFH, and thus modeled as:
 
 \begin{equation}
 SFR(t) = \begin{cases}
   \exp{-t/\tau_1} & \text{if } t < t_1 - t_2 \\
   \exp{-t/\tau_1} + k \times \exp{-t/\tau_2}      & \text{if } t \geq t_1 - t_2
  \end{cases}
\end{equation}

\noindent where $\tau_{2}$ is the e-folding time of the young stellar population, and $k$ the amplitude of the second exponential which depends on the burst strength parameter $f_{ySP}$\footnote{$f_{ySP}$ is defined as the fraction of stars formed in the second burst versus the total stellar mass formed.}.
Finally, the delayed SFH is defined as:

\begin{equation}
SFR(t) \propto t\,exp(-t/\tau_1).
\end{equation}

We assume a Salpeter IMF and the stellar population models of \cite{Maraston05}.
The metallicity is fixed to the solar one, 0.02. 
Dust extinction is  modeled assuming the \cite{Calzetti00} law with $E(B-V)_{*}$ in  the range of value shown in Table~\ref{fitparam}.  
We also assume that  old  stars  have  lower  extinction compared  to  young  stellar populations by a fixed  factor, $f_{att}=0.44$ \citep{Calzetti00}.  
The UV/optical stellar emission  absorbed  by  dust  is  remitted  in  the  IR  assuming  the \cite{Dale14} templates.  
We emphasize that this is different from the \cite{DraineLi07}  libraries  used to  generate  the mock  photometric catalogue (see Section~\ref{build} and Table~\ref{mockparam}). 
This reduces the impact on the results of using the same assumptions/templates to generate and to fit  the  photometry of  simulated  extragalactic  sources.

AGN templates are  also included in the fitting procedure.
In  Appendix~\ref{iffitnoagn}, we demonstrate  that in  the case  of AGN hosts  this  is essential  to  minimize  biases  in stellar  mass  and SFR estimates related  to contamination of the stellar light   by   AGN   emission.    
Ignoring  this   effect   results   in an over-estimation  of  the   stellar  mass  by  up  to   150\%  and  the SFR by up to  300\% depending on the spectral type and the strength of the AGN component. 
The parameters used to fit the AGN    component    are   chosen    based    on    the   results    of \cite{Fritz06}.  
However,  we decide  to  fix  the  value of  $\beta$, $\gamma$, and $\theta$, that  parametrized the density distribution of the dust within the torus, with typical values found by \cite{Fritz06} to limit  the number  of models. 
Indeed, allowing for different values of these parameters would result in degenerated model templates.

Although  the AGN parameters used to create the mock SEDs are not used in the fitting part of this work, we recognize that there could be potential problems with using the same template library to generate
mock photometric data  and then fit them.  
As a test, we have created the mock galaxies with observed AGN templates and fit them with the \cite{Fritz06} models, the results are discussed in Section~\ref{Lusso}.
We find that our results and conclusions are robust to the set of templates used to model/fit the AGN component of the SED.

The modeled SEDs are integrated  into the selected set of filters, and these  modeled flux densities  are then  compared to  the ones  of the input catalogue of  galaxies. 
For each galaxy in  the mock photometric catalogue,  and for  each set  of  model parameters,  the $\chi^2$  is computed.  
The code then  builds the probability distribution function (PDF) of  the derived  parameters of interest  (e.g.  stellar  mass, SFR, $frac_{AGN}$) based  on the  $\chi^2$ value  of the  fits.  
The output value  of a  parameter is the  mean value  of the PDF,  and the associated error  is the standard  deviation determined from  the PDF.
We  refer the  reader to  \cite{Noll09}  for more  information on  the Bayesian-like analysis performed by CIGALE.

To avoid any bias that could rise from the fact that we use the  same tool  to  create and  analyze  the mock  SEDs,  we take  the following precautions:

\begin{itemize}

	\item The flux densities of the mock SEDs are perturbed by adding a noise randomly taken in a Gaussian distribution with a standard deviation of 0.1. 
	\item We use simple analytic SFHs in the SED fitting procedure in order to reproduce \galform\ SFHs.
	\item Although the mock galaxies dust emission is modeled with the \cite{DraineLi07} library, we use the \cite{Dale14} templates in the SED fitting.
	\item Even though we also use the \cite{Fritz06} library to perform the fitting, we prevent CIGALE to use the templates used in the building of the mocks catalogues. We did not use a different AGN library for our fitting procedure for two reasons. The first one is that we are  limited by the AGN models available in CIGALE. The second one is that it has been shown in \cite{Feltre12} that the smooth torus library of \cite{Fritz06} is highly degenerated with the clumpy torus library of \cite{Nenkova08a}. Thus using for instance this very different library will not affect our results.
\end{itemize}

\begin{table*}
	\centering
	\caption{Parameter ranges used in the blind fitting procedure. }
	\begin{tabular}{l c  c }
 	\hline\hline
	Parameter & Symbol & Values\\ 
	\hline		
	\hline
	\multicolumn{3}{c}{Star Formation History} \\  
	\hline
	Metallicity  & $Z$ & 0.02 \\
	IMF & & Salpeter \\
	\hline
	\multicolumn{3}{c}{Double exponentially decreasing} \\
	\hline
	$\tau$ of old stellar population models (Gyr) & $\tau_1$ & 1, 3, 5\\
	Age of old stellar population models (Gyr) & $t_1$ & 1, 2, 3, 4, 5\\
	$\tau$ of young stellar population models (Gyr) & $\tau_2$ & 10. \\
	Age of young stellar population models (Gyr) & $t_2$ &  0.01, 0.03, 0.1, 0.3\\
	Mass fraction of young stellar population & $f_{ySP}$ & 0.001, 0.01, 0.1, 0.2\\
	\hline
	\multicolumn{3}{c}{Single exponentially decreasing} \\
	\hline
	$\tau$ of stellar population models (Gyr) & $\tau$ & 0.5, 1, 3, 5, 10\\
	Age of stellar population models (Gyr) & $t$ & 1, 2, 3, 4, 5\\
	\hline
	\multicolumn{3}{c}{Delayed SFH} \\
	\hline
	$\tau$ of stellar population models (Gyr) & $\tau$ & 0.5, 1, 3, 5, 10\\
	Age  (Gyr) & $t$ & 4, 5, 5.5\\		
	\hline
	\hline
	\multicolumn{3}{c}{Dust Attenuation} \\  
	\hline		
	Colour excess of stellar continuum light for the young population &$E(B-V)_*$ & 0.05, 0.1,0.15, 0.2, 0.25,  0.3,0.35, 0.4, 0.5,0.6	 \\
	$E(B-V)_*$ reduction factor between old and young populations & $f_{att}$ & 0.44 \\
	\hline		
	\hline
	\multicolumn{3}{c}{Dust template} \\  
	\hline
	IR power-law slope & $\alpha$		& 1.5, 2, 2.5	\\
	\hline		
	\hline
	\multicolumn{3}{c}{AGN emission} \\  
	\hline
	Ratio of dust torus radii &$R_{max}/R_{min}$ &  30, 100 \\
	&$\tau_{9.7}$ &   0.3, 3.0, 6.0, 10.0\\	
	&$\beta$ &   -0.5\\		
	&$\gamma$ &  0.00 \\
	&$\psi$ &   0.001, 50.100, 89.990\\
	Opening angle of the torus & $\theta$ &  100 \\
	Fraction of $L_{IR}$ due to the AGN\tablefootmark{a} & $frac_{AGN}$ & -0.2, -0.15, -0.1, -0.05, 0.0, 0.05, 0.1, 0.15, 0.2\\ 
	  &   & 0.25, 0.3, 0.4, 0.5, 0.6, 0.7, 0.8\\ 
	\hline
	\label{fitparam}
	\end{tabular}
	\tablefoot{
		\tablefoottext{a}{We use low negative values of $frac_{AGN}$ in order to minimize a bias due to PDF analysis, as explained in Section~\ref{agncon}.}
	}
\end{table*}

\begin{figure}%[!h] 
	\includegraphics[width=\columnwidth]{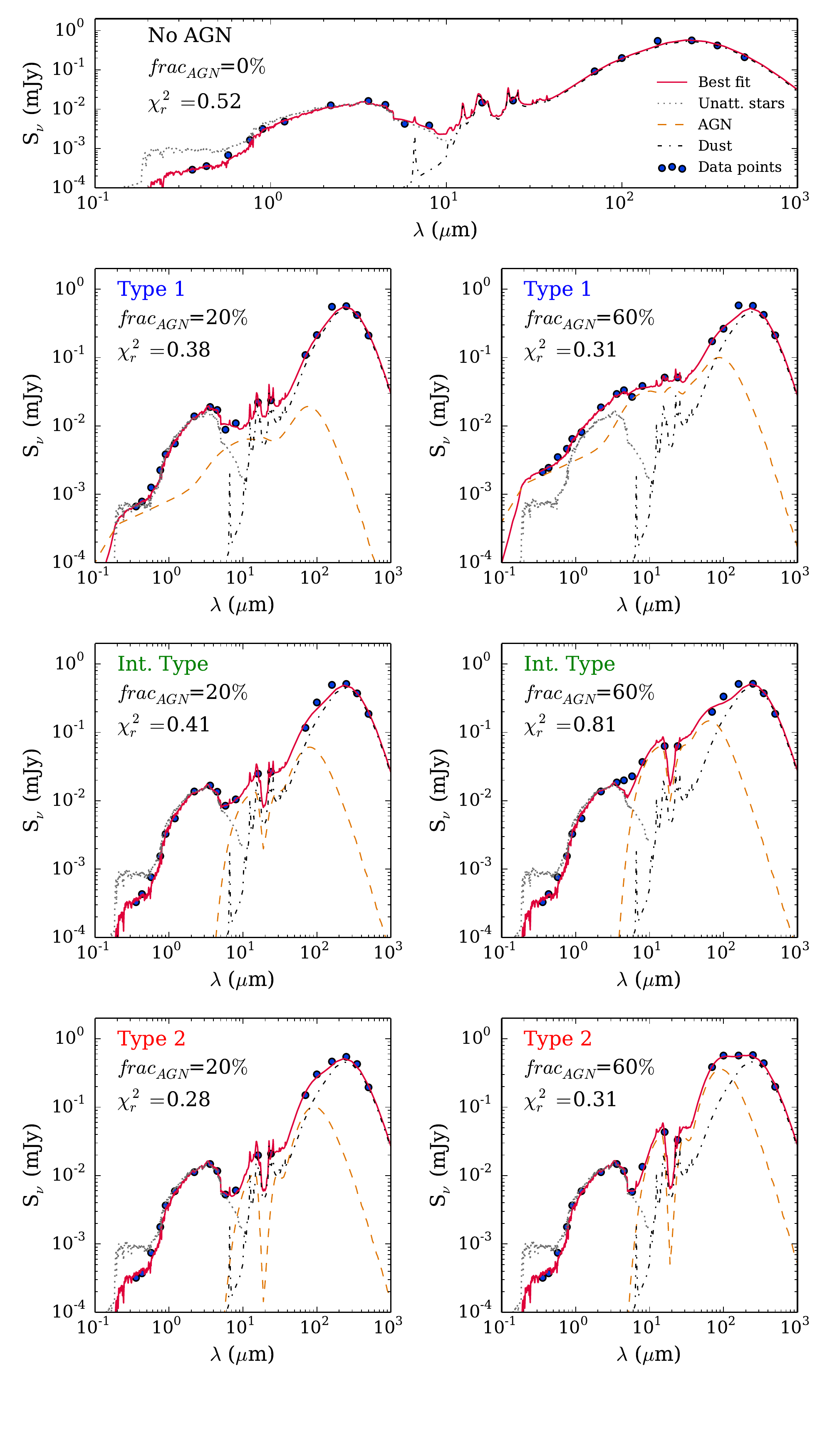}
	\caption{ \label{bestSEDs} Examples of fits for mock galaxies  associated with one of the SFHs shown is Figure~\ref{sfh_plot}. Blue points are the flux densities of the mock galaxies, the red lines are the best fits obtained by CIGALE. In addition, we show the unattenuated stellar emission in black dotted line, the dust emission in black dashed-dotted line, and the total emission from the AGN in orange dashed line. The associated reduced $\chi^2$ are provided. The single top panel shows the best fit in the absence of AGN emission. The two upper panels show the best fit for Type 1 with a fraction of 20\% (left) and a fraction of 60\% (right), the middle ones for the intermediate type, and the bottom panels for the Type 2 AGN.  } 
\end{figure}

We present in Figure~\ref{bestSEDs} examples of fits of the mock galaxies corresponding to the \galform\ SFH presented in Figure~\ref{sfh_plot}, and assuming a double exponentially decreasing SFH.	
In Type 1 and intermediate type AGNs, the fits are good but we note some difficulties in reproducing measurement near the IR peak. 
This disagreement is attributed to compatibility problems between \cite{Dale14} and \cite{DraineLi07} dust emission libraries in this range \citep{Ciesla14} that have no impact on the results of this study.
For the intermediate type AGN, Figure~\ref{bestSEDs} shows that the code uses a model with a strong silicate absorption at 9.7\microns.
However, as shown in Figure~\ref{inputsed} (middle panel), the input AGN model used for the intermediate type does not show any silicate absorption.
Thus, despite the availability of models with low values of $\tau_{9.7}$ in the fitting procedure, the code does not reproduce the absence of silicate absorption of the input SED.
Although these points have no impact on the result of this study, they give a glimpse on possible problems to constrain AGN templates with broad-band photometry.

\begin{figure*}%[!h] 
	\includegraphics[width=\textwidth]{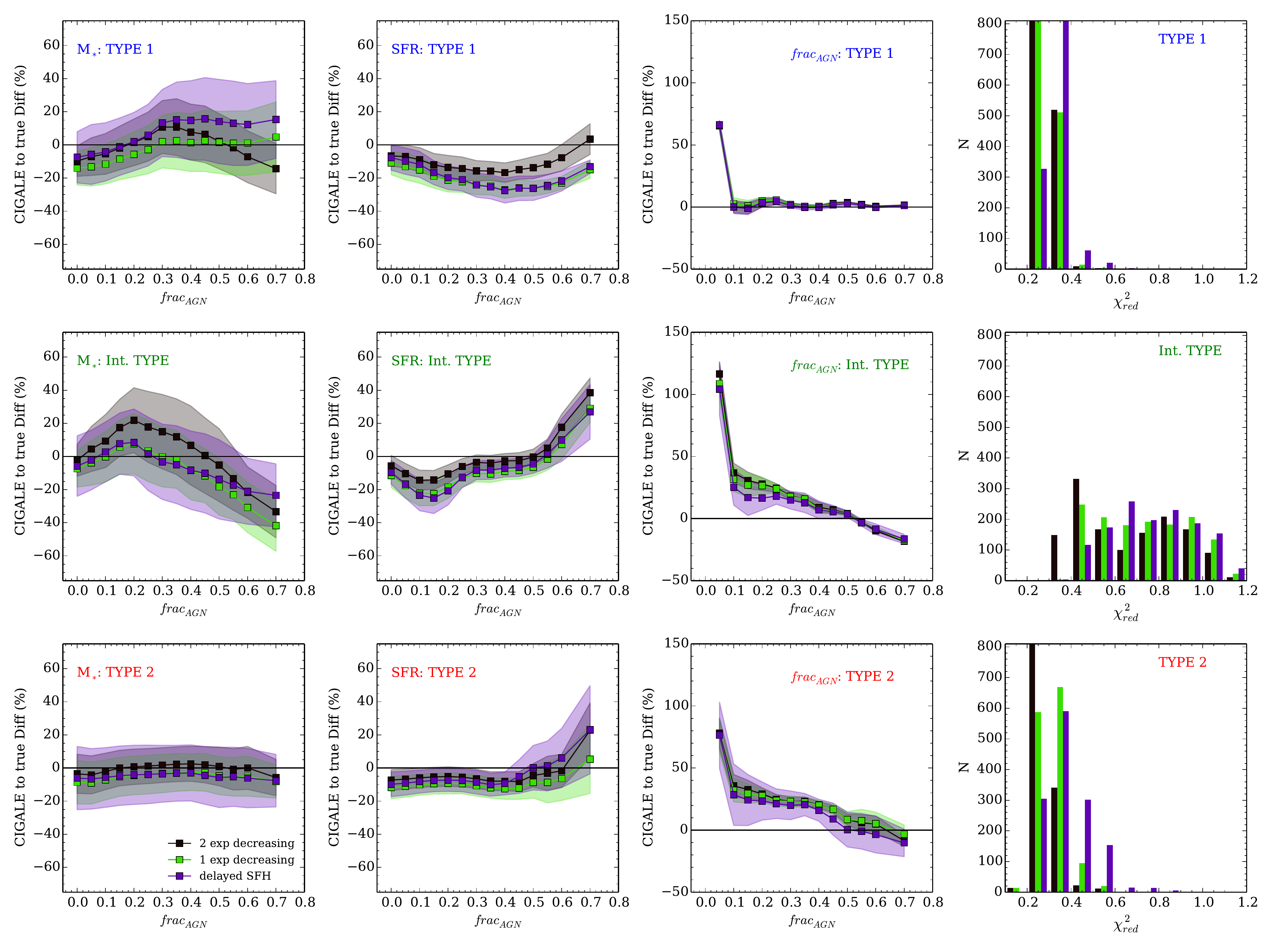}
 	\caption{ \label{sfhs} Fractional    difference   between parameters derived from CIGALE and the simulated ones as a function of $frac_{AGN}$, i.e. the contribution of the AGN light to the overall IR luminosity. The first column of panels plots the fractional difference in  stellar mass estimates.  The second  column is  for SFR estimates and the  third column corresponds to $frac_{AGN}$.  The fourth column of panels shows the distribution of $\chi_{red}^2$. Each row corresponds  to a  set of simulation  that assume  different input SEDs for the AGN component. From  top to bottom, we present results for Type 1  AGN, intermediate  type and Type  2 AGN. Different  colors in each panel correspond to different functional forms for the SFH. Black is for the 2$\tau$-dec model, green corresponds to the 1$\tau$-dec model, and  purple marks the delayed  SFH. Dots represent the mean value of the fractional difference at each input $frac_{AGN}$, the shaded regions show the one $\sigma$ scatter.}
\end{figure*}

The estimates of the $M_*$, SFR, and $frac_{AGN}$ as a function of the contribution of the AGN are presented in Figure~\ref{sfhs} for the three different assumptions made on the SFH, as well as the $\chi^2_{red}$ distribution for each case.
We discuss these results in the absence of AGN in Section~\ref{noagn}, and the impact of the AGN contribution in Section~\ref{withagn}.

%=================================================================================	
	\subsection{\label{noagn}Determining stellar masses and star formation rates in simulated galaxies without AGN component}

	\begin{table}
		\centering
		\caption{Fractional differences between the derived $M_*$ and SFR  and the input ones for the mock galaxies with $frac_{AGN}=0$ (no AGN contribution). }
		\begin{tabular}{c c c c c }
	 	\hline\hline
		SFH model &   \multicolumn{2}{c}{$M_*$} & \multicolumn{2}{c}{SFR}\\ 
							&mean (\%) & $\sigma$ (\%) & mean (\%) & $\sigma$ (\%)\\ 
		\hline
		1$\tau$-dec 	& -10.0 & 11.5 & -11.4 & 7.0\\
		\hline		
		2$\tau$-dec 	& -5.2  & 10.2& -6.6 & 6.2\\
		\hline
		delayed 		& -6.5& 17.6 & -9.1& 7.4\\
		\hline
		\label{offsets}
		\end{tabular}
	\end{table}

	\begin{figure}%[!h] 
		\includegraphics[width=\columnwidth]{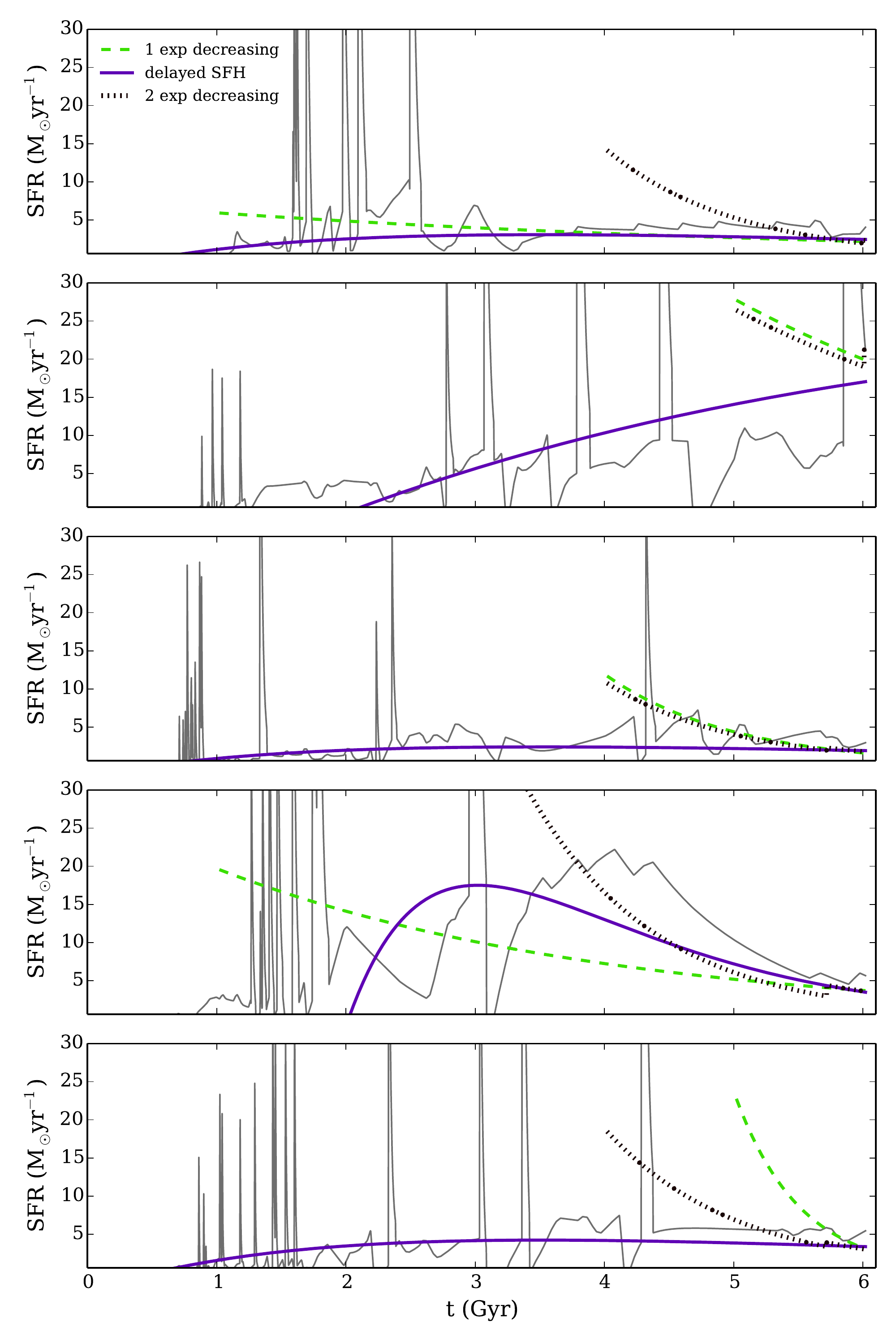}
  		\caption{ \label{compsfh} Comparison between five SFHs from \galform\ and the corresponding best-fit output SFHs obtained by CIGALE. The green dashed-line is for the 1$\tau$-dec model, the purple solid-line corresponds to the delayed SFH, and the black dotted-line to the 2$\tau$-dec model.}
	\end{figure}
	
	We first explore fractional  differences between the derived and input stellar masses  and SFRs for  mock galaxies without any  AGN component included   in   their  SEDs,   i.e.    the   case  $frac_{AGN}=0$   in 	Figure~\ref{sfhs}.  
	Both parameters depend  on the adopted SFH used to fit the  mock galaxy multi-waveband  photometry.  
	Different functional forms for the SFH are expected to induce systematic variations in both $M_*$  and SFR  determinations \citep{Bell03,Lee09,Maraston10,Pforr12}.
	Table~\ref{offsets}  provides  the fractional  systematic offsets for the different SFH assumptions used in our analysis to fit the mock galaxy photometry.   
	The model that provides  on average the best agreement between the output  and input $M_*$ and SFRs is the 2$\tau$-dec.  

	In order to compare the output parameters linked to SFH, we show on Figure~\ref{compsfh} five \galform\ SFHs as well as the associated SFHs obtained from the best fit of the SED fitting procedure.
	\galform\ SFHs are complex and analytical SFHs cannot reproduce the numerous star formation bursts.
	However, as the integral of the SFH provide the stellar mass of the galaxy, it is clear that, in order to recover the stellar mass, the best 1$\tau$-dec and 2$\tau$-dec models use a very small, unrealistic age.
	Indeed, if we fix the age of the galaxy, contrary to what is usually done in the literature when using these assumptions, by providing a smaller age range, between 4 and 5.5\,Gyr for our samples as galaxies are at $z$=1, then these two models overestimate $M_*$ by 15 and 19\%, respectively.
	From the three SFHs models used in this work, the delayed SFH better reproduces the global envelop of the \galform\ SFH.
	As a  result this functional form yields stellar masses in reasonable agreement with the input ones and, at  the same time, a galaxy  formation age close to  the simulated one.
	This is consistent with the mean SFHs of \textsc{Illustris} presented in \cite{Sparre14}.
	Indeed, the mean and median SFHs of \textsc{Illustris} sources have shapes that can be typically modeled with a delayed SFH.
	This result obtained from Semi-Analytical Models of galaxy evolution seems to be in agreement with the conclusions of \cite{Boselli01} based on observations of local galaxies.
	We thus conclude that a delayed SFH seems to be a more realistic assumption on the SFH of galaxies.

	Several published works have studied the ability of SED fitting techniques to retrieve the stellar mass of galaxies.
	They were based on mock catalogues, on SAMs, or on hydrodynamical codes, and some of them made use of the IR domain \citep{Wuyts09,Lee09,Pforr12,Pacifici12,Mitchell13,Buat14}.
	Also using \galform\ SFHs, \cite{Mitchell13} showed that a single exponentially decreasing SFH provides a good estimation of $M_*$ with a small offset of -0.03\,dex.
	Using CIGALE, \citep{Buat14} found that the output stellar mass is systematically lower by 0.07\,dex compared to the true one.
	Despite the different methods used in these works, our results are in good agreement as we find an underestimation of the stellar mass of $\sim$7\% averaged over the three types of SFH.

	Small differences are also found in the derivation of the SFR from the different SFH assumptions, but they globally slightly underestimate the SFR with offsets between 6.6 and 11.4\% (Table~\ref{offsets}).
	These relatively good estimations ($<$12\%) obtained with the SFR are in perfect agreement with the results of \cite{Buat14} who found that the SFR is robustly estimated, with systematic differences lower than 10\%, whatever the chosen SFH model as long as one IR data is available.

	Another parameter which is known to be directly linked to the estimation of the stellar masses and SFR is the amount of attenuation, quantified in this work by $E(B-V)_*$.
	The link between the attenuation and the stellar mass is however indirect. 
	Without a strong constraint on the dust attenuation, a degeneracy between the age of the old stellar population (which is directly linked to the stellar mass) and the attenuation appears \citep{Pforr12,Conroy13,Buat14}.
	We  explore this  bias  by generating  mock galaxy  photometry  by varying  the $E(B-V)_*$  input value and  fixing $frac_{AGN}$=0.  
	The resulting  mocks catalogues are fit  with a  2$\tau$-dec  SFH  model.  
	The  systematic  offset of  the derived $M_*$ increases by a factor of two between $E(B-V)_*=0.05$ and $E(B-V)_*=0.7$. 
	The  underestimation on the  SFR varies from  -15\% to -6\%   for    $E(B-V)_*=0.05$   ($A_V$=0.2\,mag)   to   $E(B-V)_*=0.5$ ($A_V$=2\,mag).   
	Furthermore, if we use, in the mock catalogues, a host galaxy SED with a completely obscured star formation, then the contribution of the AGN to the total IR luminosity is still recovered for high fractions, but the offset between 15$\%$ and 30$\%$ is larger by a factor of $\sim$2.
	Thus the  level  of stellar  light attenuation  also plays a  role in recovering the  stellar mass and SFR  of galaxies, as already shown in previous studies \citep[e.g.,][]{Wuyts09,Mitchell13,Buat14}.
	
	In conclusion, the 2$\tau$-dec model provides the best estimates of $M_*$ and  SFR of the simulated galaxies, in the expense of unrealistic galaxy ages.
	The delayed SFH recovers the stellar mass with a mean offset of $\sim$6.5\%, and a mean offset on the SFR of $\sim$9\%, but better reproduces the true SFH of the galaxies. 
	We use the results obtained in absence of AGN emission as references to analyze the AGN impact on the estimation of $M_*$ and SFR.
%%%%%%%	
	\subsection{\label{withagn}Determining stellar masses and star formation rates in AGN host galaxies}
	
	As shown in Figure~\ref{fluxratios}, depending on its intensity, the AGN emission contaminates large parts of the SED especially the key domains used to retrieve the stellar mass and the SFR of the underlying galaxy.
	In this Section, we discuss the ability of broad band SED fitting to constrain this contamination, decompose  the AGN from  the host  galaxy light,  determine  stellar masses  and SFR of  AGN  hosts.
	The results are presented in Figure~\ref{sfhs} for the three AGN types considered in this work, where we show the mean fractional difference for each parameter as a function of the input power of the AGN (points), as well the 1-$\sigma$ scatter (shaded regions).
	
		\subsubsection{\label{agncon}Constraining the AGN contribution}
		
\begin{figure*}%[!h] 
			\includegraphics[width=6cm]{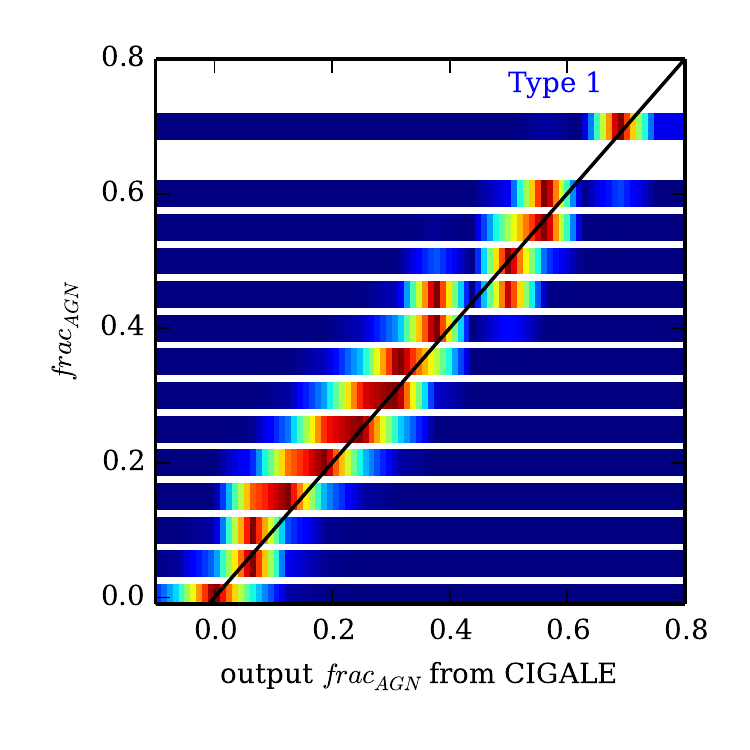}\hfill
			\includegraphics[width=6cm]{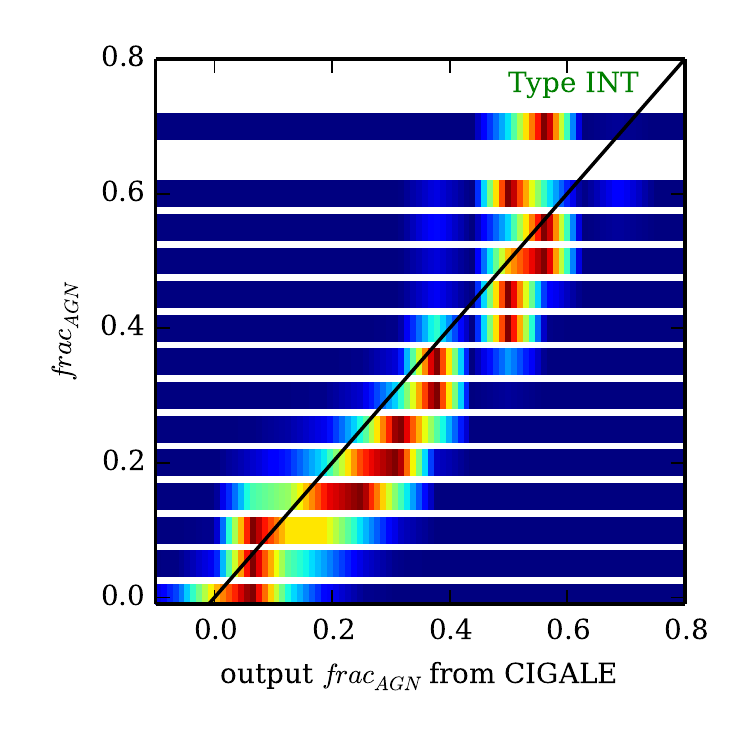}\hfill
			\includegraphics[width=6cm]{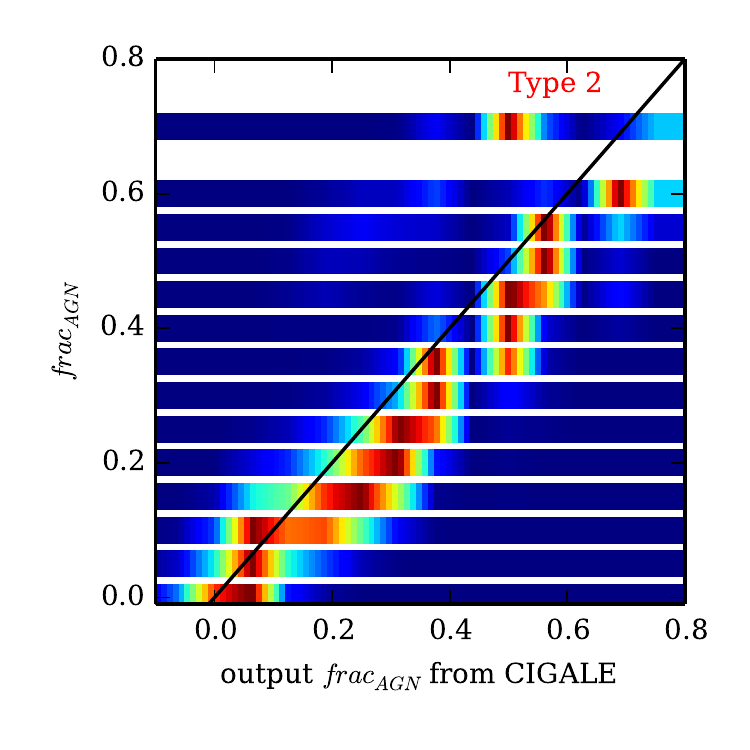}\\
	  		\caption{ \label{PDFfracAGN} Probability distribution function for the $frac_{AGN}$ determination of a  particular \galform\ SFH.  The vertical axis in  all  panels corresponds  to  different input $frac_{AGN}$. Each  panel corresponds to a different AGN SED added to the stellar light: left is for the Type 1  AGN, the middle panel corresponds  to the intermediate AGN type,  and right panel plots  the results for the  Type 2 AGN. The color  indicates  the level  of  probability,  blue  being the  lowest probability  and  red  the  highest. The black solid line is the one to one relationship.}
		\end{figure*}

		First we explore the  ability of CIGALE  to constrain  the fractional contribution of  AGN emission  to the total  IR luminosity,  i.e.  the $frac_{AGN}$  parameter.  
		The  set of  panels in  the third  column of Figure~\ref{sfhs} presents the relative difference between the output and input  $frac_{AGN}$ averaged  over  the 100  \galform\  SFHs, for  the different  assumptions  of  SFHs. 
		The AGN contribution is almost always overestimated for all three types, except for high fraction values ($frac_{AGN}>$50\%) for the intermediate type and Type 2 AGNs, where it is underestimated.
		Independent from the input AGN  SED shape (Type 1, intermediate type, or Type 2), below $frac_{AGN}$=10\%, there is a large overestimation of the AGN contribution, that can reach up to $\sim$120\%.
		One explanation could be the well-known effect of the use of PDF analysis in retrieving parameters.
		For the lowest value of the parameter, the PDF will be truncated, and thus taking its mean value will slightly shift the output value of the parameter toward a larger value, yielding to its overestimation, as explained in \cite{Noll09} and \cite{Buat12}.
		However, to prevent this effect, we provide low negative $frac_{AGN}$ as well as input parameters.
		Without any impact on the results of the SED fitting and the estimates of $M_*$ and SFR, it allows a better estimate of very low $frac_{AGN}$ from the PDF analysis.
		Thus, the overestimation observed for very low fractions shows the difficulty CIGALE has to distinguish between a SED without any AGN contamination and an SED with a very low contribution of the AGN ($\sim$5\%).
		We thus conclude that low AGN contributions are very difficult to constraint from broad band SED fitting with an overestimation up to a factor of 2 for $frac_{AGN}<$10\%.
		At higher fractions, the overestimation depends on the type of AGN.
		A Type 1 contribution to the total $L_{IR}$ is well recovered with an offset smaller than 10\% for fractions higher than 10\%.
		In order to better understand this trend, we show in Figure~\ref{PDFfracAGN} (left panel) the PDF for each $frac_{AGN}$ as seen from the top of it, for the SEDs associated to one SFH.
		We can see from Figure~\ref{PDFfracAGN} that the shape and position of the PDF do not change when $frac_{AGN}$ increases.
		For the intermediate type and Type 2, an overestimation of 30-40\% can be expected for $frac_{AGN}$ between 10 and 40\%.
		Then the offset decreases at higher fractions.
		These effects are seen on Figure~\ref{PDFfracAGN}, the PDF is large for low values of $frac_{AGN}$ and shows a constant overestimation up to high values of $frac_{AGN}$, i.e. 60\%.		
				
		\subsubsection{Estimating the stellar mass}
		
		The  panels  in  the   first  column  of  Figure~\ref{sfhs}  plot  the fractional difference between the input  stellar mass of AGN hosts and that  inferred from  the template  fits to  the mock  photometry. 
		This figure shows that the stellar masses  of AGN hosts can be derived with systematic  uncertainties  smaller  than 40\% ($\Delta  \log   M_*< 0.15$\,dex),   even  in   the   case   of  Type 1  AGNs and up to $frac_{AGN}=0.7$.  
		This underlines the importance of AGN/stellar light decomposition  when fitting  templates to  multi-wavelength photometric data of  AGN to  derive properties of  the underlying galaxy.   
		If AGN templates were not included  in the  analysis  the inferred  stellar masses would  be biased to  high values by  a factor as large as 2.5 (see Appendix~\ref{iffitnoagn}).
	
\begin{figure*}%[!h] 
			\includegraphics[width=6cm]{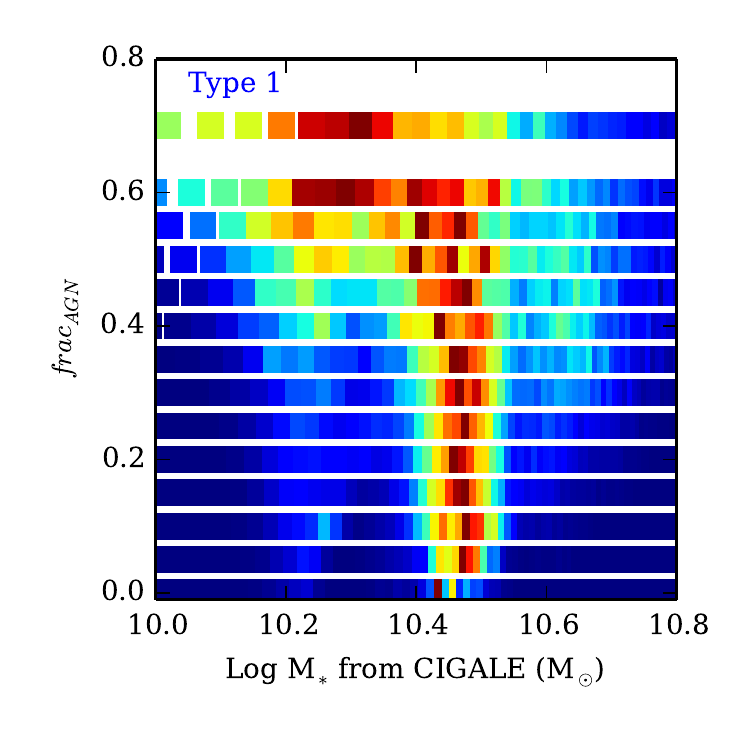}\hfill
			\includegraphics[width=6cm]{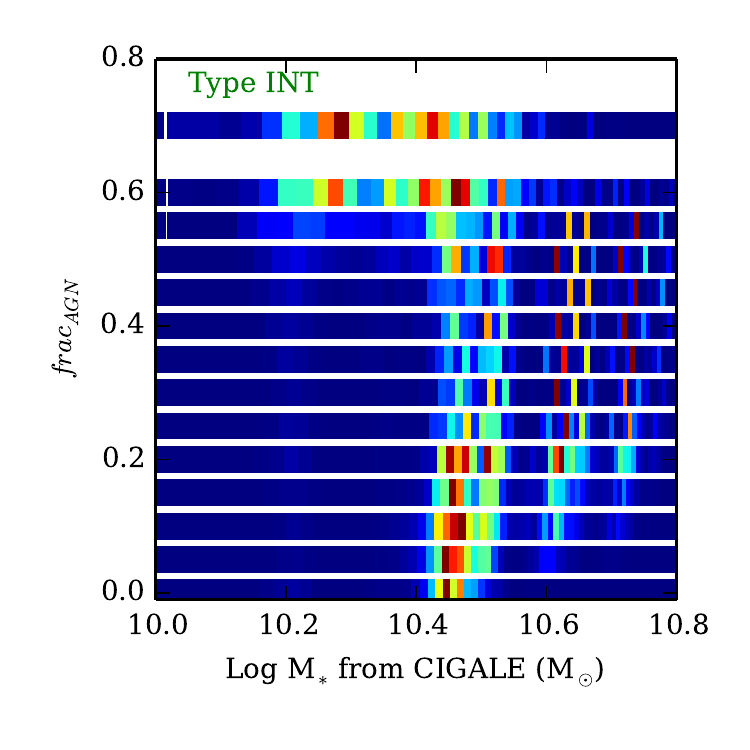}\hfill
			\includegraphics[width=6cm]{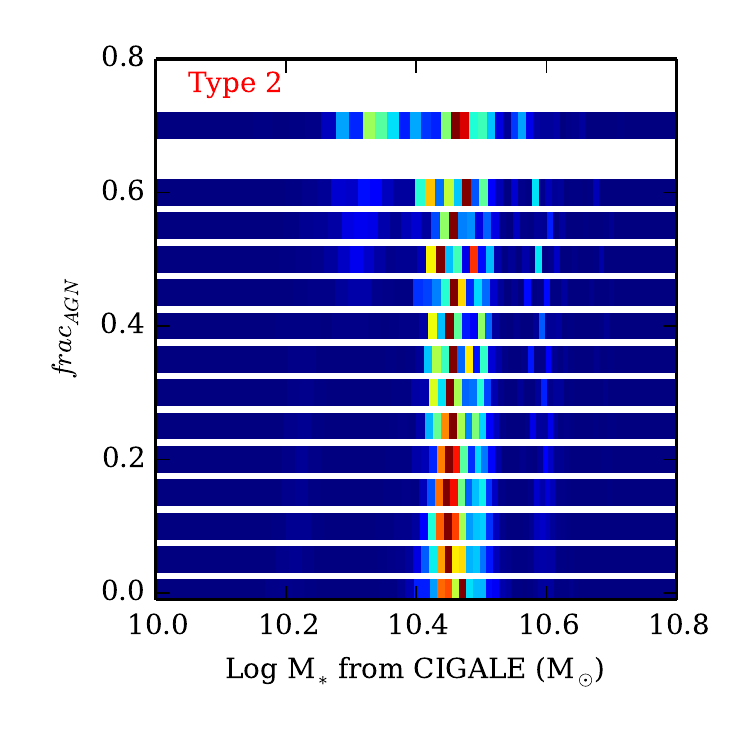}\\
 	 		\caption{ \label{PDFMstar} Probability distribution function for the $M_*$ determination of a  particular \galform\ SFH.  The vertical axis in  all  panels corresponds  to  different input $frac_{AGN}$. Each  panel corresponds to a different AGN SED added to the stellar light: left is for the Type 1  AGN, the middle panel corresponds  to the intermediate AGN type,  and right panel plots  the results for the  Type 2 AGN. The color  indicates  the level  of  probability,  blue  being the  lowest probability  and  red  the  highest. The true $\log  M_*$  for  the particular  simulated  galaxy is  10.39  in  solar  units.}
		\end{figure*}

		Careful inspection of Figure~\ref{sfhs} also suggests that, in the case of Type 1 AGN, the uncertainties on the estimation of $M_*$ as a function of Type 1 AGN fraction seem to not depend on the assumption made on the SFH up to $frac_{AGN}=$0.4.
		The fractional difference shows a constant increase up to $frac_{AGN}<$0.4 and then reaches a plateau for higher fractions, except for the 2$\tau$-dec model for which the offset decreases.
		Indeed, as we can see from Figure~\ref{fluxratios} (upper panel), at an AGN fraction of 20\%, the AGN emission is higher than the host emission by a factor of 2 in the end of the NIR domain.
		As NIR flux densities are known to be a proxy for the stellar mass as the emission is dominated by the old stellar population \citep[e.g.,][]{Gavazzi96}, it is sensible to think that there is a link between the contribution of the AGN in NIR and the variations observed in Figure~\ref{sfhs}.
		To understand this trend, we see in Figure~\ref{PDFMstar} (left panel) the PDF of the stellar mass for each $frac_{AGN}$ as seen from the top of it.
		The PDF is very broad and is skewed toward lower values of $M_*$ when the $frac_{AGN}$ increases.
		The variation of the fractional difference in Type 1 AGNs depends on the contribution of the AGN to the total $L_{IR}$.
		A very weak effect can be attributed to the assumption made on the SFH but the 1$\sigma$ scatter of each of them are overlapping showing that this effect is marginal.
		
		A systematic effect is seen in the intermediate AGN type (Figure~\ref{sfhs}, first column, middle panel).
		The offset on the stellar mass slightly increases with the contribution of AGN up to $frac_{AGN}\sim$20--30\% and then decreases.
		This threshold of $frac_{AGN}\sim$20--30\% corresponds to the point where the $frac_{AGN}$ starts to be relatively well constrained, even if it is still overestimated.
		We note that the delayed SFH appears relatively less affected showing weaker amplitude of variations with $frac_{AGN}$.
		Given the same pattern observed for three SFHs considered, the variation seen here is mostly due to the AGN emission.
	
		For Type 2 sources, the AGN emission has no influence on the derivation of $M_*$ whatever the SFH chosen. 
		This is likely due to the high obscuration of the light emitted from the accreting supermassive black hole, affecting only slightly the observed rest-frame NIR flux of the galaxy.
		Indeed, it is clear from Figure~\ref{fluxratios} that the NIR rest frame domain is not contaminated by the AGN emission.
		The $M_*$ PDF is relatively narrow and shows the offset discussed in Section~\ref{noagn} (Figure~\ref{PDFMstar}, right panel).

		\subsubsection{Estimating the star formation rate}
\begin{figure*}%[!h] 
			\includegraphics[width=6cm]{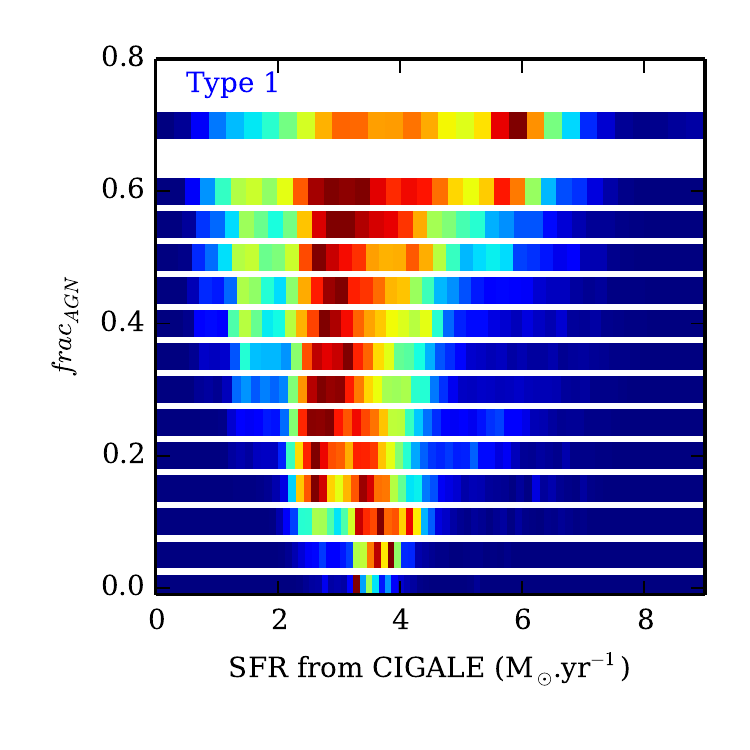}\hfill
			\includegraphics[width=6cm]{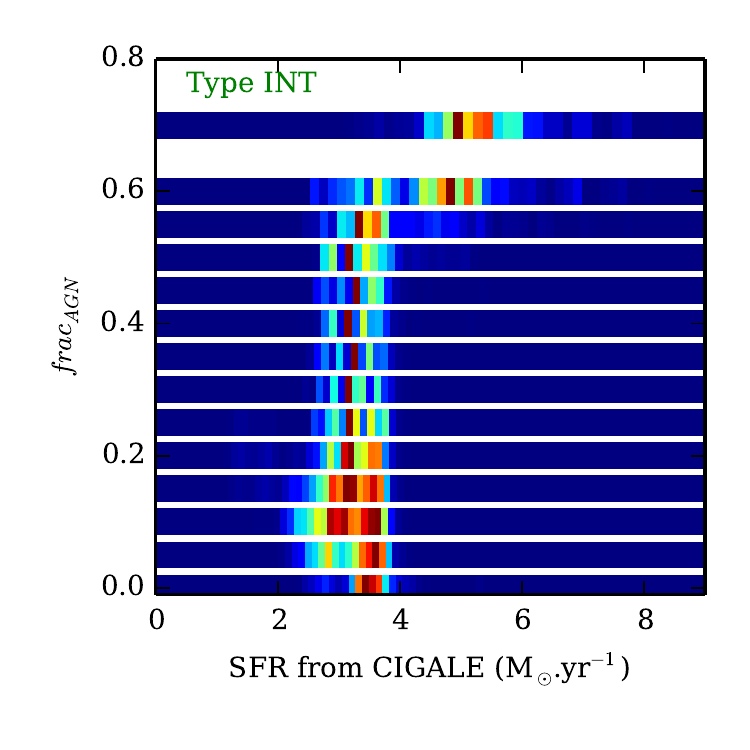}\hfill
			\includegraphics[width=6cm]{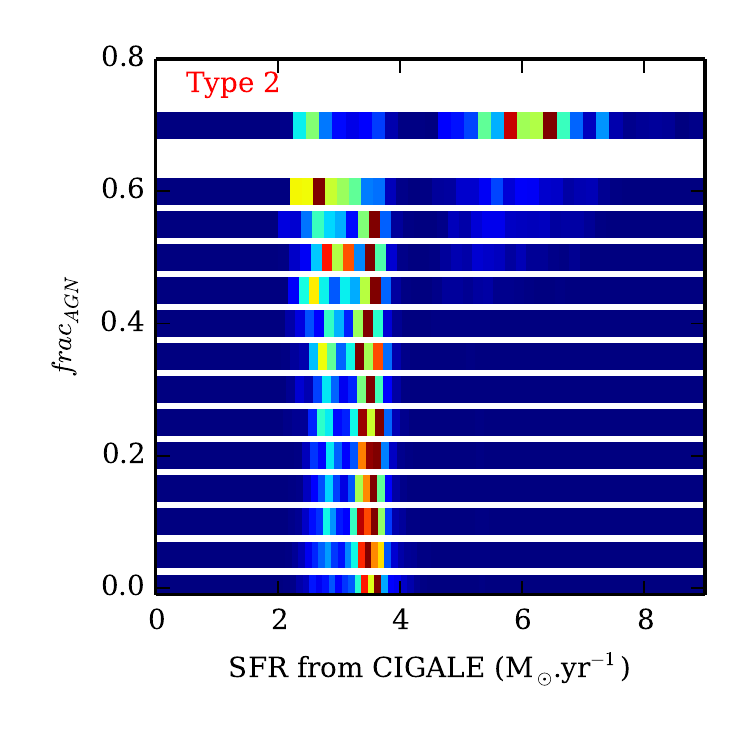}\\
	  		\caption{ \label{PDFSFR} Probability distribution function for the SFR determination of a  particular \galform\ SFH.  The vertical axis in  all  panels corresponds  to  different input $frac_{AGN}$. Each  panel corresponds to a different AGN SED added to the stellar light: left is for the Type 1  AGN, the middle panel corresponds  to the intermediate AGN type,  and right panel plots  the results for the  Type 2 AGN. The color  indicates  the level  of  probability,  blue  being the  lowest probability  and  red  the  highest. The true SFR  for  the particular  simulated  galaxy is  3.3\,M$_{\odot}$.yr$^{-1}$.}
		\end{figure*}

		For all three types of AGN considered in this work, the variation of the SFR estimation as a function of AGN strength is the same for the 1$\tau$-dec, 2$\tau$-dec, and delayed SFH.
		This implies that these variations are entirely due to the AGN emission.
		For Type 1 AGNs, the underestimation of the SFR increases with $frac_{AGN}$ up to 50\%.
		Indeed, as shown on Figure~\ref{fluxratios}, the UV becomes totally dominated by the AGN emission very quickly.
		However, the FIR is not much affected by the AGN emission which only dominates the 30\microns\ rest frame emission by a factor of 3 for the highest $frac_{AGN}$.
		This allows us to still constrain the attenuation and thus the SFR, even when there is an AGN contamination and estimate the SFR with a maximum offset of 30\%.
		The example of Figure~\ref{PDFSFR} (left panel) shows that the PDF is close to the true value, but in this case, a secondary peak arises for AGN fractions higher than 35\%, yielding to the under-estimation.
		
		In the  case of the intermediate AGN type  and for  input AGN fractions  of $frac_{AGN}=$10\%  the host galaxy SFR is  underestimated by a factor of  20-30\%.  
		This is likely related  to the  overestimation of  the  AGN contribution  to the  FIR luminosity  (see Figure~\ref{sfhs} third  column, middle  panel). 
		This results  in an  underestimation of  the  host galaxy  emission in  the MIR-FIR domain, and thus an  underestimation of the SFR.  
		However, for input $frac_{AGN}$  in the interval 25--50\%, the  AGN contribution to the FIR luminosity is better constrained. 
		As a result the inferred SFR from the SED fit is also  in better agreement with the input one.  
		For input   $frac_{AGN}\ga50$\%  the  opposite   is  happening.   
		The  AGN contribution to  the FIR is  underestimated and hence the  host galaxy emission   in   the   rest-frame   MIR-FIR   part  of   the   SED   is overestimated. 
		The net effect is  an overall overestimation of the SFR up   to    40\%.     
		This   behavior    is    also   shown    in Figure~\ref{PDFSFR} (middle  panel) where the  peak of the PDF is  slightly shifted toward lower values  up to $frac_{AGN}$=30\%, then lies  on the  right values up  to $frac_{AGN}$=60\% where  the SFR starts to be overestimated for  very high fractions.

		For Type 2 AGNs, we find the same pattern as in the two previous cases.
		The 1$\tau$-dec, 2$\tau$-dec, and delayed SFH follow the same trend.
		The estimation of the SFR is quite insensitive to the AGN contribution with a very weak decreasing up to $frac_{AGN}$=40\%.
		For higher fractions, the offset on the SFR measurement increases up to an over-estimation up to 20\% for the 2$\tau$-dec and the delayed SFH models.
		A comparison between the SED with a AGN contribution of 40\% and the SED of normal galaxy shows an emission three times higher for the AGN SED in the FIR domain, mandatory to have a good estimation of the SFR \citep{Buat14}. 
		The SFR PDF corresponding to the 2$\tau$-dec model (Figure~\ref{PDFSFR}, right panel) shows two peaks, one on the right value and one slightly overestimating the SFR.
		For high AGN fractions, a third peak arises overestimating even more the SFR and skewing the PDF toward higher values.
		However, this increase is weak for the 1$\tau$-dec SFH model yielding to overestimation of 5\%--10\% at $frac_{AGN}$=70\%.

%=================================================================================
	\subsection{Impact of the photometric coverage}
	
	The results presented in Section~\ref{noagn} and Section~\ref{withagn} have  been  tested  in  the  ideal  case  where  complete  photometric coverage, from UV  to submm rest frame, is  available for each source.
	In  this  section,  we study  how  the  lack  of photometric  data  at different  spectral bands  affects  the determination  of the  stellar mass,  the SFR,  and the  $frac_{AGN}$ parameter.   
	As shown in  Section~\ref{noagn}, there is no perfect  SFH assumption, we thus arbitrarily  choose the 2$\tau$-dec SFH model.   
	This choice does not affect the  discussion because we are interested  in deviations of the parameters inferred from spectral fits using incomplete broad-band coverage relative to the ideal case that all spectral bands listed in Table~\ref{filt} are available. 
	We run our SED fitting procedure using exactly the same parameters as in  the previous sections. 
	However, in  successive runs,  the  input  mock   catalogues  lack  photometry  in  certain broad-band  filters.  
	We  explore  in particular  how  the results  on stellar  mass,  SFR, and $frac_{AGN}$  change  if  we exclude in turn, the sub-mm rest frame ( \textit{Herschel}/SPIRE 350\microns\ and 500\microns),  the FIR rest frame (filters  with $\lambda_{mean}\ge24$\microns\  in Table~\ref{filt}), the mid-IR rest frame  (filters with  $\lambda_{mean}\ge8$\microns\ in Table~\ref{filt}), and the UV rest frame (filters with  $\lambda_{mean}<0.762$\microns\ in Table~\ref{filt}).
	
\begin{figure*}%[!h] 
		\includegraphics[width=\textwidth]{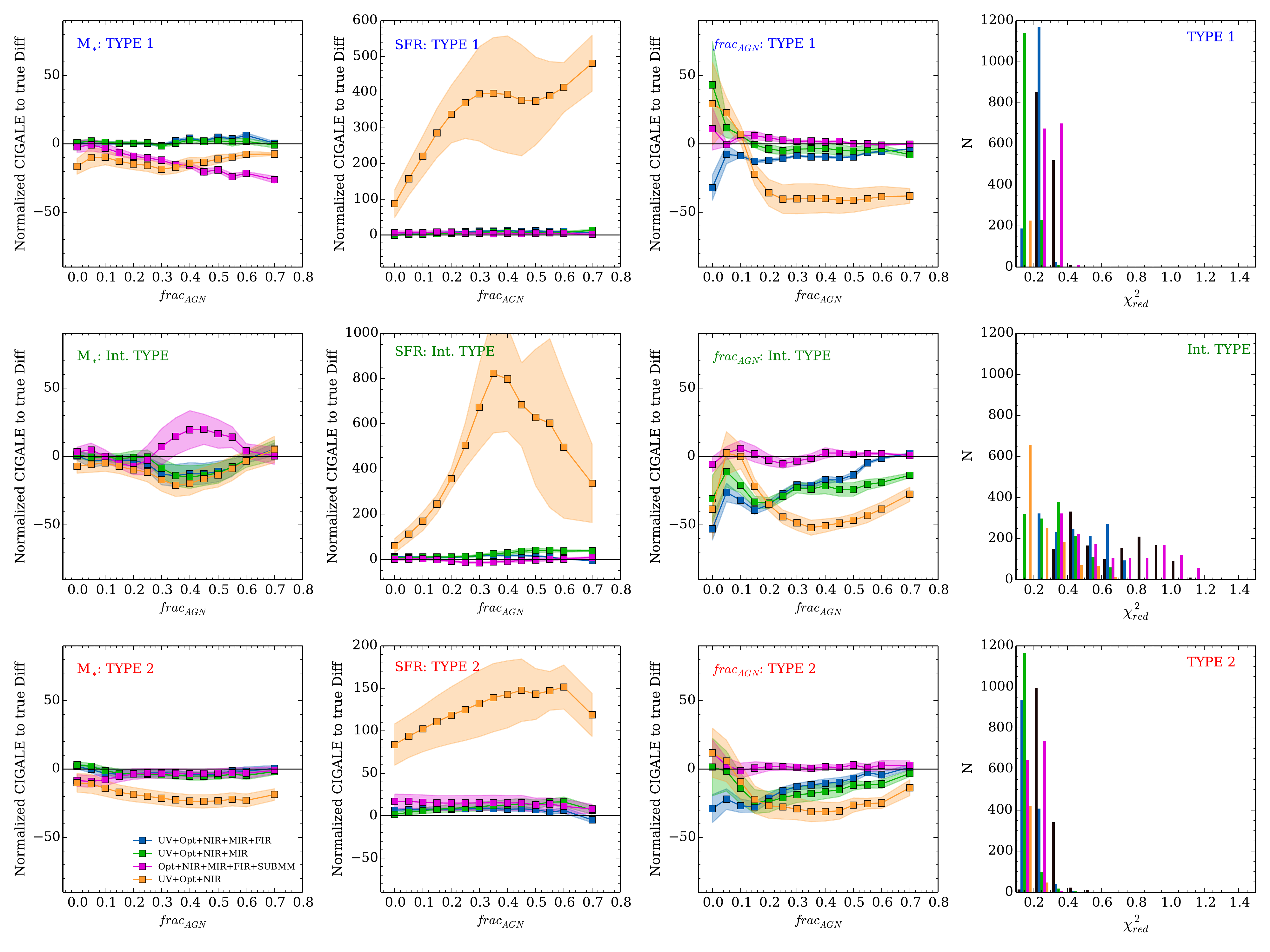}
  		\caption{ \label{mstarsfrvsfracAGNbandes} Evolution of the relative difference between output parameters from CIGALE and the true ones with the input fraction of AGN for the three types of AGN considered, normalized to the values obtained with a full photometric coverage. The blue relation corresponds to a photometric coverage containing no submm photometry. Green corresponds to the absence of FIR-submm data. Orange corresponds to the absence of IR data. Finally, the UV was removed for the magenta relation. $\chi^2_{red}$ distribution are shown for every run on the fourth column, we add the distribution corresponding to the total photometric coverage in black for comparison.}
	\end{figure*}

	Figure~\ref{mstarsfrvsfracAGNbandes}  presents  the  results for  each trial.
	The impact of the different spectral ranges in constraining the AGN contribution is shown in the third column of Figure~\ref{mstarsfrvsfracAGNbandes}.
	The lack  of UV rest frame photometry has  no impact in  the determination of  $frac_{AGN}$, even for Type 1 AGNs.  
	This parameter is more sensitive to the availability of FIR and submm photometry.  
	For all input AGN spectral types and for input  $frac_{AGN}\ga10-20\%$,  the  lack of  IR  and/or sub-mm  data results in a systematic underestimation of inferred $frac_{AGN}$ up to typically 20-30\% 		relative to the case  of complete photometric coverage.  
	Intermediate  type and  Type2 AGN show  larger deviations compared to Type 1 AGNs.  
	Interestingly, for input $frac_{AGN}\la10-20\%$, the  largest offset is  found in  the absence  of submm  data, showing the  importance  for  these  long  wavelengths  in  constraining  the emission   from   dust   	heated   by   young   and   evolved   stellar populations.  
	The exclusion of  all photometric  bands above 8\microns\ produces   systematic  offsets   of  up   to  50\%   in   the  derived $frac_{AGN}$. 
	This is  expected because CIGALE is based on energy balance  and  therefore  requires  long wavelength  data  for  optimum performance \citep{Noll09}.	

	The first column  of panels in Figure~\ref{mstarsfrvsfracAGNbandes} shows changes  in the  determination of  stellar mass  for different  set of photometric bands.  
	The absence of FIR and/or submm has no significant impact on  the derivation  of the  stellar mass for  any of  three AGN spectral types.  
	The  observed systematic variations with $frac_{AGN}$ are  similar  to  the  case  of complete  photometric  band  coverage.
	However, we  note a small  underestimation of $\sim$10\% in  $M_*$ for intermediate  type with input  $frac_{AGN}$ between  30 and  60\%. 
	The lack of UV photometry has no  impact on the determination of $M_*$ for Type 2 AGN. 
	This  is because in this case there is  no AGN emission to contaminate  the  UV--to--NIR bands.   
	The  absence  of UV  photometry however, does affect  the $M_*$ estimates for Type 1 and intermediate type AGN. 
	It leads to an underestimation of the stellar mass in Type 1 AGNs,  and  an overestimation  for  intermediate  type  AGN for  input $frac_{AGN}=30 - 60\%$.   
	This underlines the importance of  the UV to recover  the stellar  mass  for these  types  of AGNs.   
	When no  data longward  of observed  8\microns\  are  available, the  stellar mass  is always  moderately underestimated by  up to  20-30\%, relative  to the case of  full photometric  band coverage. 
	This  is in  agreement with \cite{Noll09} who found an underestimation of 0.19\,dex in the case of local normal (i.e. no AGN component) galaxies for the same combination of photometric data.

	The   panels  in   the  second   column  of Figure~\ref{mstarsfrvsfracAGNbandes}   shows   that  the   photometric coverage has  an impact on the  determination of SFR  in AGN dominated
	galaxies.  
	The strongest effects are observed in absence of data above observed 8\microns.  
	This  yields large  systematic  errors of 150-800\% in SFR,  depending on the level of  input $frac_{AGN}$. 
	This is in agreement with \cite{Noll09} who also found an overestimation of the SFR by  0.83\,dex for normal galaxies (i.e. no AGN  component) for the same photometric data coverage. 
	Smaller  offsets in SFR estimates are found when the FIR, sub-mm and UV photometric bands are excluded. 
	For Type 1 and Type 2 AGNs, the absence of any of these spectral domains leads to a systematic  overestimation of 5--20\%.  
	In the  case of intermediate type  AGNs,  the removal  of  the submm  leads  to  variations in  the estimation of the  SFR up to $\sim$20\%.  
	The  situation is worse when no  FIR data  is  available, where  the  offset on  the SFR  increases rapidly with  input $frac_{AGN}$ up to  35\%.  
	The absence  of UV data for intermediate type AGNs has a peculiar effect  as it starts to have an impact from  $frac_{AGN}$=20\%,  where  the   SFR  is  underestimated  up  to $frac_{AGN}$=60\%.  
	In  intermediate types, the  AGN has no  impact in the    UV   domain,    as   shown    in    Figure~\ref{inputsed}   and Figure~\ref{fluxratios}.  
	Removing this range forces CIGALE to rely on the  FIR  to  estimate the  SFR,  which  is  contaminated by  the  AGN emission.  
	It  is therefore  difficult to constrain  the SFR  of objects with  AGN  SED  components   similar  to  the  intermediate  type  AGN considered in this work without any UV data.
	
	Regarding constraints on the stellar  mass in the case of SEDs without AGN contribution ($frac_{AGN}=0$), very small differences, less than a few  percent, are  observed  when reducing  the available  photometric bands.  
	This  is expected since  $M_*$ is mainly constrained  from the NIR rest frame  emission.  
	The estimation of the SFR  differs by up to $\sim$10\% when  we remove  IR and  UV data.  
	This  is to  be expected since these spectral domains are important to constrain the SFR.
		
%=================================================================================
	\subsection{\label{Lusso} Impact of the AGN library}	

\begin{figure*}%[!h] 
		\includegraphics[width=\textwidth]{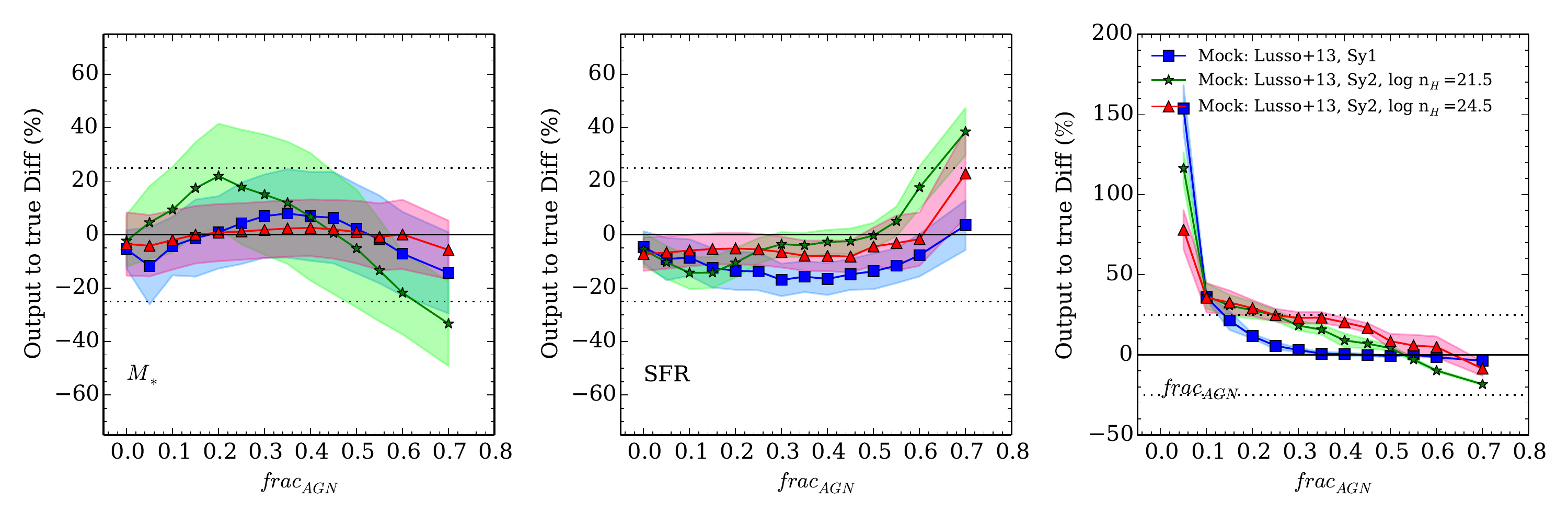}
 		\caption{ \label{lusso} Comparison between the output values of $M_*$, SFR, and $frac_{AGN}$ and the true ones in the case where the AGN emission of the mock galaxies is modeled using \cite{Lusso13} templates. The SED fitting is made using the 2$\tau$-dec SFH models, and the \cite{Fritz06} models. The results for the Type 1 sample are in blue, for the intermediate type sample are in green, and for the Type 2 sample in red.}
	\end{figure*}

	In order to focus our study on the biases that AGN emission has on the estimate  on  $M_*$  and  SFR,  we modeled  the  mock  galaxies  using templates from  the \cite{Fritz06} library, and used  the same 	library to perform  the SED fitting. 
	Although  we are cautious not  to use the same  AGN  templates  when  simulating  and fitting  the  mock  galaxy photometry, it may be possible that the trends we observe are at least partially driven by 		degeneracies among  the broad-band AGN SEDs of the \cite{Fritz06} library.  
	We  explore this by using a  different set of AGN  templates to  build the  mock galaxy  photometric  catalogue.  
	We select three SEDs presented  by \cite{Lusso13}, derived by \cite{Silva04}, their Seyfert 1, their mildly  obscured Seyfert  2 with  $\log N_H$=21.5\,$\rm  cm^{-2}$, and their heavily obscured Seyfert  2 with $\log N_H$=24.5\,$\rm cm^{-2}$.
	These templates bear similarities to the the Type 1, Intermediate type, and  Type 2  \cite{Fritz06} AGN templates  defined in  Table~\ref{mockparam}.   
	The approach  described  in  Section~\ref{sedmod}  is  followed  to  construct  mock galaxy/AGN  composite photometry,  with the  only difference  that the three \cite{Lusso13} templates above  are used instead of  the \cite{Fritz06} ones.  
	The SED fitting of the resulting mock galaxy catalogues follows the steps described in Section~\ref{sedmod} \citep[i.e. using ][templates]{Fritz06}. 
	Figure~\ref{lusso} presents  in the same way  as Figure~\ref{sfhs} the results on  stellar mass, SFR, and $frac_{AGN}$  for the 2$\tau$-dec SFH  model.  
	For  these parameters the  overall systematic trends  in  Figure~\ref{lusso}  are   similar  to  those  observed  in Figure~\ref{sfhs}  where  the  modeling  of the  mock  galaxies  is  made with \cite{Fritz06} models.   
	We are  therefore confident that  our results are insensitive to the adopted AGN template library. 
	
%=================================================================================
\section{SFR--$M_*$ relation \label{ms}}

\begin{figure*}%[!h] 
	\includegraphics[width=6cm]{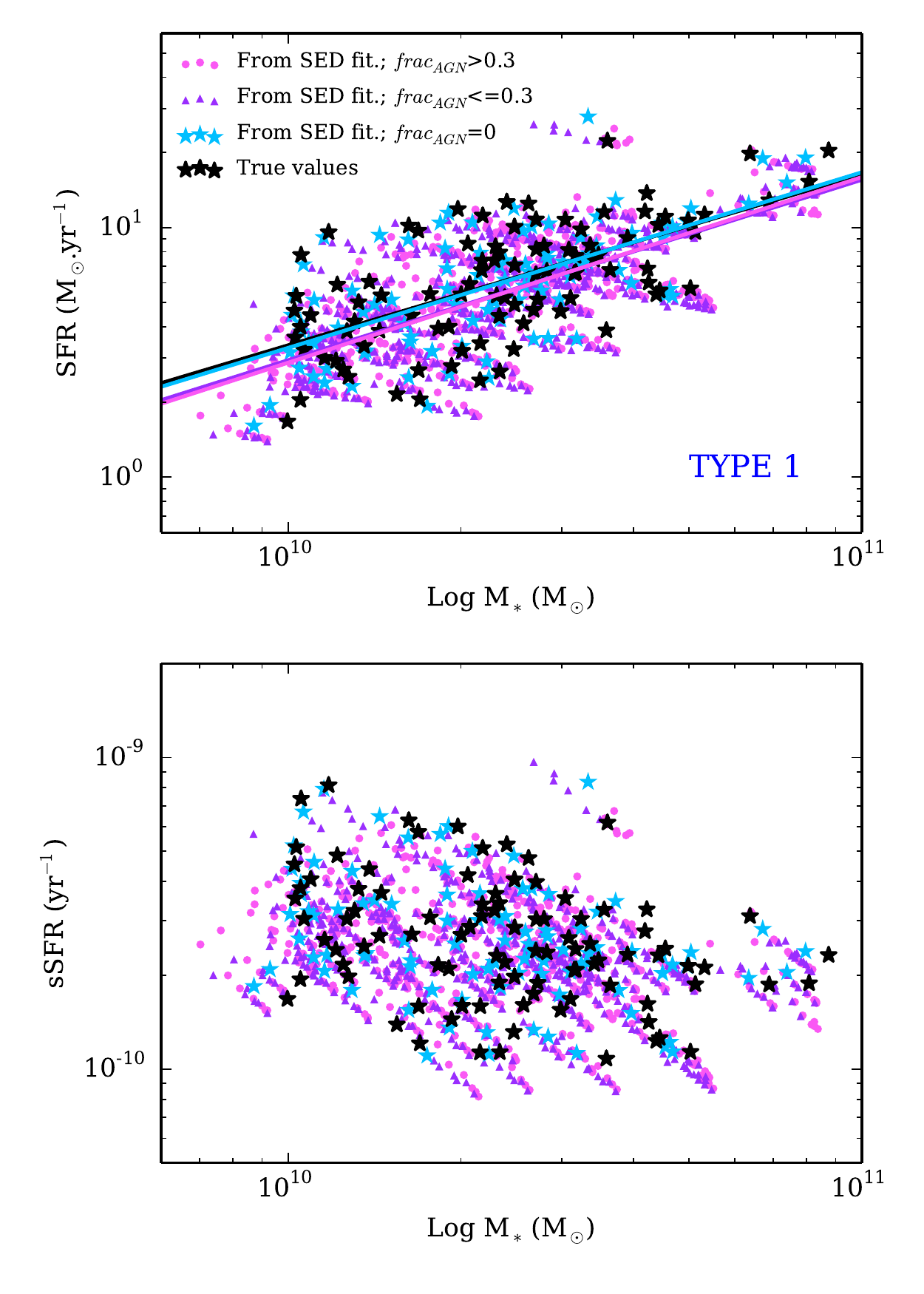}
	\includegraphics[width=6cm]{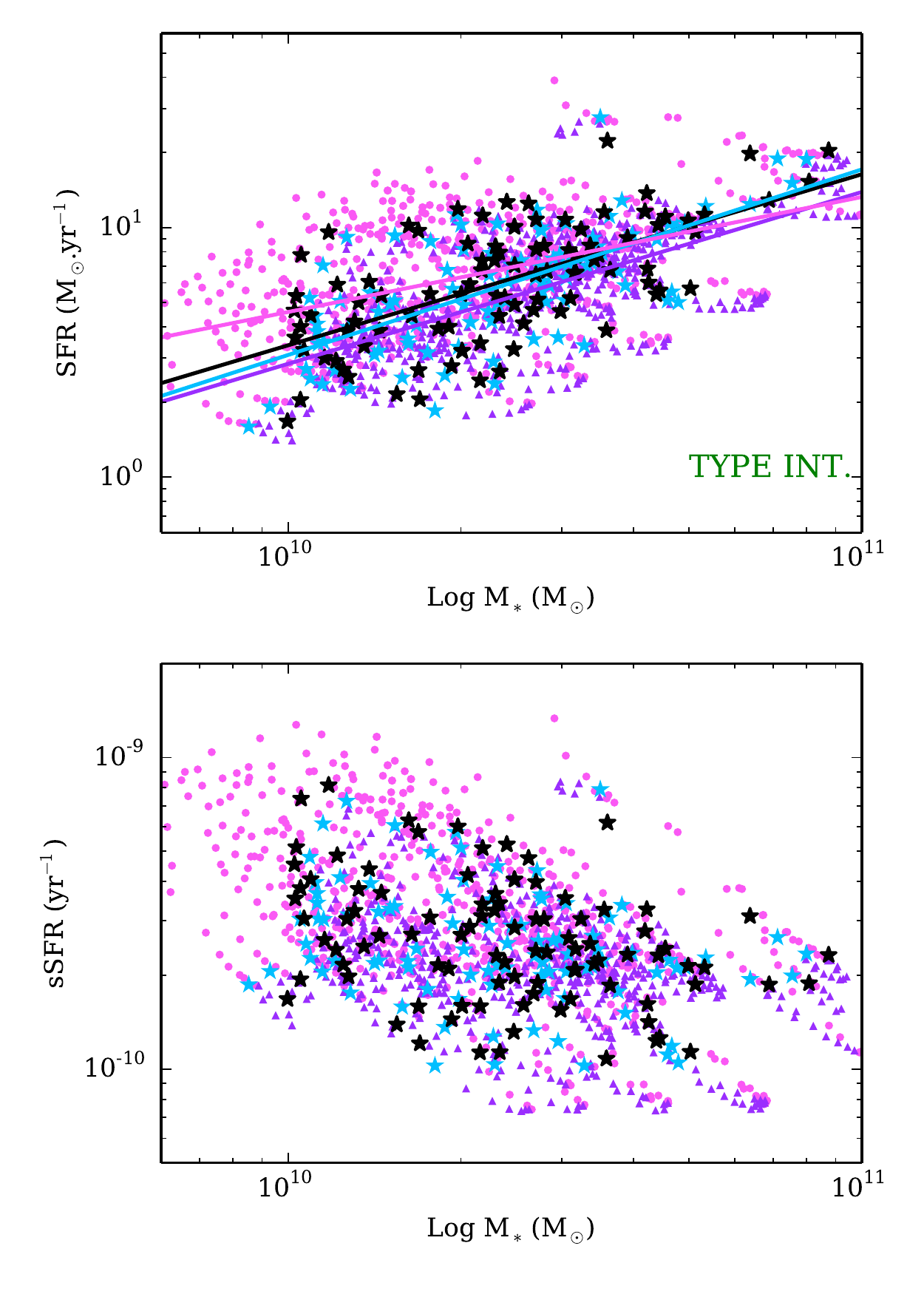}
	\includegraphics[width=6cm]{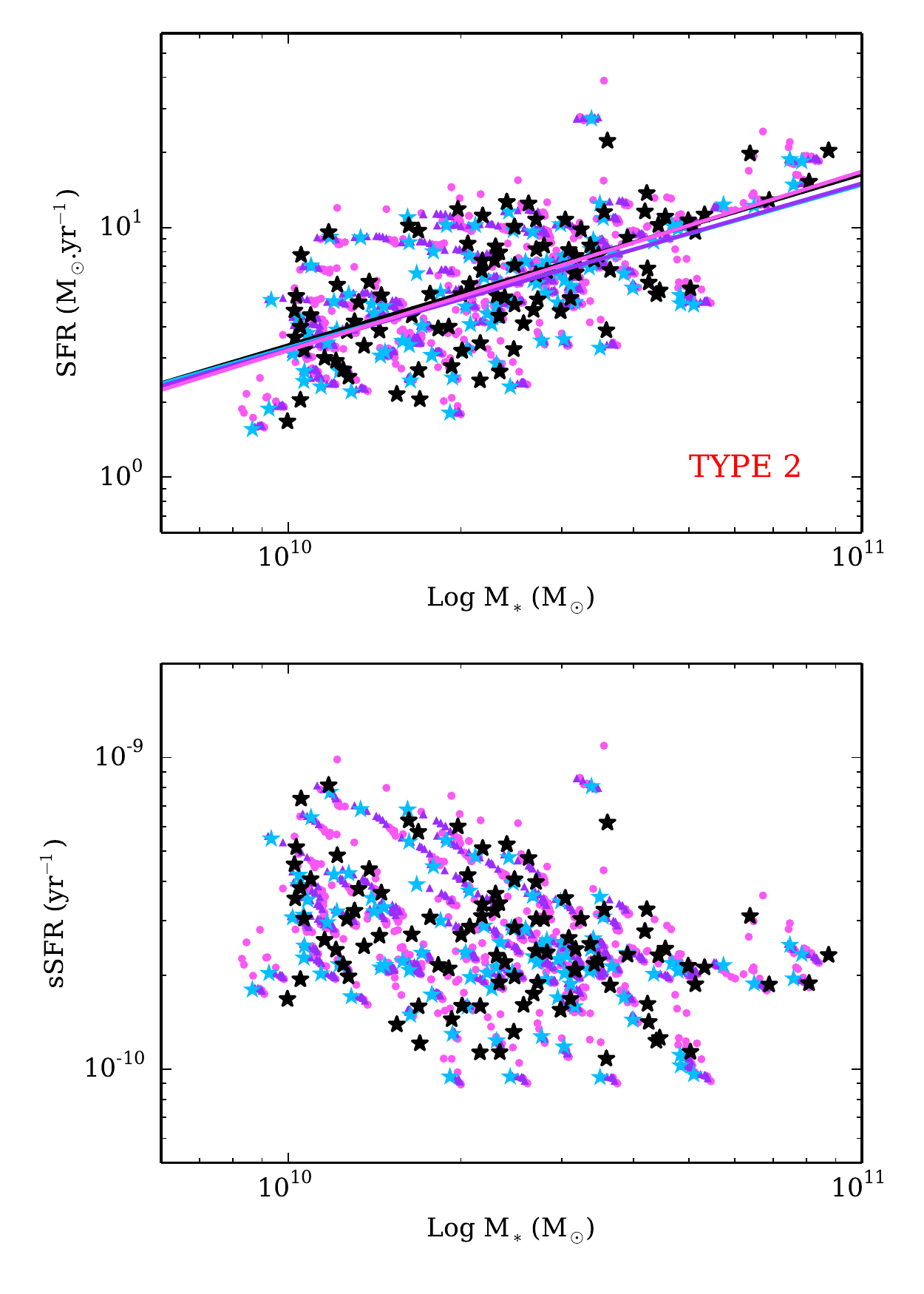}\\
 	\caption{ \label{sfrmass_comp}  Top panels: SFR-$M_*$ diagram for the Type 1 sample (left), the intermediate type (middle), and the Type 2 sample (right) obtained when using the 2$\tau$-dec SFH model for the fitting. Black stars are the true values of the SFR and $M_*$ of the mock galaxies. Cyan stars are the galaxies without any AGN component ($frac_{AGN}$=0), purple triangle are galaxies with an AGN contribution lower than 30\%, and pink circle galaxies with an AGN contribution higher than 30\%. Colored-lines are the linear fit corresponding to each subsample. Bottom panels: sSFR-$M_*$ diagram for the Type 1 sample (left), the intermediate type (middle), and the Type 2 sample (right). }
\end{figure*}

As mentioned earlier, SED fitting is the most popular method to derive the physical properties of very large galaxy samples often produced by deep wide area extragalactic surveys.
Recent studies of such large samples showed that the majority of star-forming galaxies are known to follow a SFR--$M_*$ correlation, called the main sequence (MS), and galaxies that lie above this sequence are experiencing a starburst event \citep{Elbaz11}.
In order to understand if the results discussed in this work could impact the shape of the MS, we show, in Figure~\ref{sfrmass_comp}, the mock galaxies of our samples in a SFR--$M_*$ plot. 
It is known that the main sequence predicted by models currently suffers normalization problems compared to the observed relation \citep[e.g.,][]{Mitchell14,Furlong14}.
At $z$=1 the MS predicted by \galform\ is about a factor of 2 lower than the observed MS \citep[see][ for a complete discussion of this issue]{Mitchell14}.
The SFHs provided by \galform\ are thus probably not a perfect representation of the real $z$=1 star forming galaxies, and it is unclear if it could affect our results.
Indeed, all the models currently available suffer from the same issue.
However, in this Section, we focus on the effect of the AGN emission and the use of SED fitting to retrieve the physical parameters of the galaxies on the MS.

As we discuss in Section~\ref{noagn} and Section~\ref{withagn}, the choice of the SFH assumption has an impact of $M_*$ and SFR, and thus on the SFR--$M_*$ correlation, as also discussed in \cite{Buat14}.
The stellar masses and SFR used in Figure~\ref{sfrmass_comp} are those obtained from a 2$\tau$-dec SFH.	
The SFR--$M_*$ relation obtained from the Type 1 sample shows a small increase of the dispersion slightly shifted toward higher stellar masses, due to the offset on the stellar mass obtained for high fractions of AGN discussed in Section~\ref{noagn} (Figure~\ref{sfrmass_comp}, upper left panel).
Thus, the increasing offset on the stellar mass estimation for Type 1 AGNs has a small impact of the SFR-$M_*$ relation.
For the intermediate type, a shift due to the stellar mass offset is observed, but in addition there is an increase of the MS dispersion due to the variation of the offsets with the AGN contribution. 
Low fractions shift the relation toward higher $M_*$ whereas the higher fractions toward lower masses, thus increasing the dispersion of the relation.
However, the Type 2 SFR--$M_*$ relation does not show a significant change as the $M_*$ estimation is totally independent from $frac_{AGN}$.
Furthermore, the underestimation observed for low AGN fraction in the estimation of the SFR seems to not affect the MS of the Type 2 sample.

The lower panels of Figure~\ref{sfrmass_comp} present the same results but for the specific SFR (sSFR).
In contrast with the Type 1 and Type 2 samples, the intermediate type sample displays a correlation between the fractional difference between the mock and output parameter values and $frac_{AGN}$.
The sample without AGN is biased toward low sSFR and higher stellar mass, and increase toward higher sSFR and lower stellar mass with $frac_{AGN}$.
The Type 1 sample only shows an underestimation of the sSFR, as well as the Type 2 sample where the underestimation is however weaker.
		
As these results depend on the assumption made on the SFH, we provide in Table~\ref{sfrmass} the linear coefficients of the best fits of the SFR-$M_*$ such as:
\begin{equation}
	\log SFR = a \times \log M_* + b.
\end{equation}
We provide these coefficients for the ``true" sample, for the sample containing no AGN, and for the Type 1, Int. Type, and Type 2 AGN samples, in all of the three SFH cases.	
In no case were we able to recover the true slope of the MS, but the closest value is obtained for the 2$\tau$-dec SFH (0.70 when the true slope is 0.68) which is the SFH assumptions providing the best estimation of both $M_*$ and SFR.
The largest the offset on the stellar mass estimation, the largest the difference to the true MS slope.
It seems that the SFR-$M_*$ relation is more sensitive to the estimation of the stellar mass than of the SFR.
The type of AGN has an influence on the MS slope for all of the three SFHs with a difference up to $\sim$32\%.
	
\begin{table}
	\centering
	\caption{Variations on the SFR-$M_*$ correlation depending on the SFH. }
	\begin{tabular}{c  c  c  c c c  }
 	\hline\hline
	 Coefficients & True & No AGN &Type 1 & Int. Type & Type 2\\ 
	\hline
	\multicolumn{6}{c}{1$\tau$-dec} \\
	\hline
	%$\sigma$ &  - & &   &   &     \\
	$a$ &  0.68 & 0.62&  0.72 & 0.47  &  0.63   \\
	$b$ &  -6.32 & -5.66  & -6.84  & -4.10 & -5.85  \\
	\hline
	 \multicolumn{6}{c}{2$\tau$-dec} \\
	\hline
	%$\sigma$ &  - & &   &   &     \\
	$a$ & 0.68 & 0.70 & 0.74  & 0.66  &   0.81  \\
	$b$ &  -6.32 & -6.51   & -6.91   & -6.11   & -7.73 \\
	\hline		
	\multicolumn{6}{c}{delayed} \\
	\hline
	%$\sigma$ &  - & &   &   &     \\
	$a$ &  0.68 & 0.77& 0.56  &  0.59 & 0.74    \\
	$b$ &  -6.32 &-7.16 & -5.16  &  -5.37 &  -6.89   \\	
	\hline
	\label{sfrmass}
	\end{tabular}
\end{table}
	
%=================================================================================
\section{Comparison against IR-only SED decomposition \label{DecompIR}}

\begin{figure*}%[!h] 
	\includegraphics[width=\textwidth]{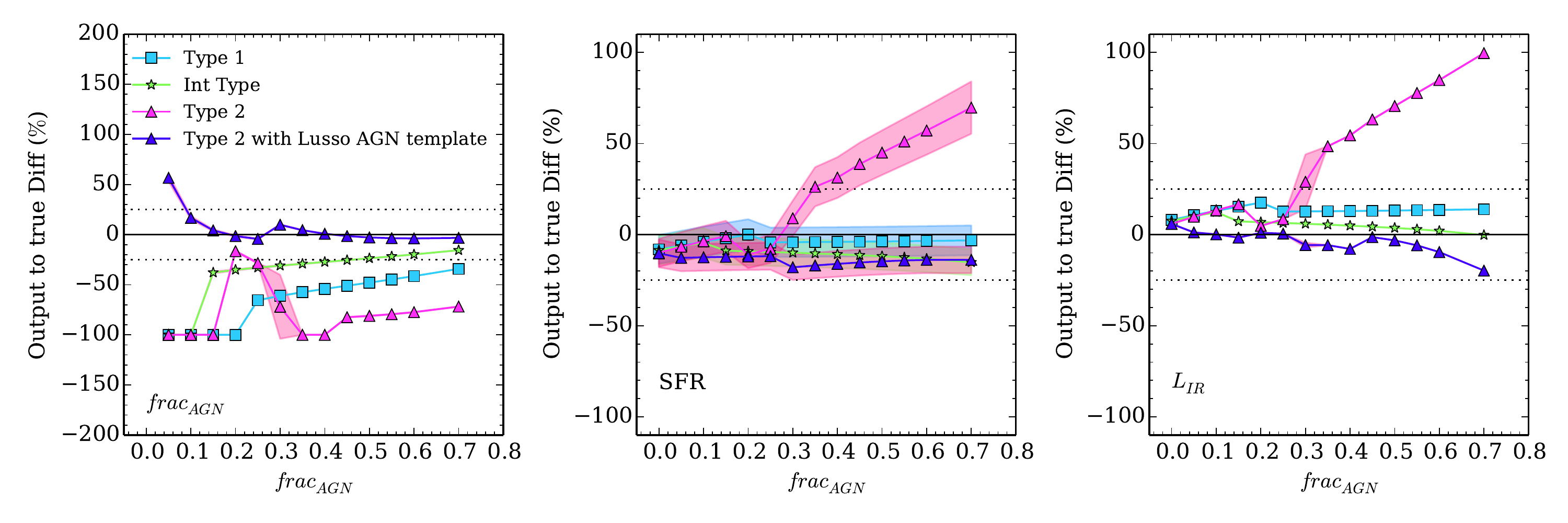}
 	\caption{ \label{decompir} Relative difference between the $frac_{AGN}$ (left panel) and the SFR (right panel) obtained by \textsc{DecompIR} and the true values. Cyan squares, green triangles, and magenta triangles are Type 1, intermediate type, and Type 2 outputs from \textsc{DecompIR}, respectively. The purple triangles present the results provided by \textsc{DecompIR} for the Type 2 mock catalogue obtained using the \cite{Lusso13} templates. The dotted lines indicate the 25\% difference.}
\end{figure*}
	
In this Section, we use our mock catalogues to test another method used to disentangle the AGN emission from the host galaxy in the IR, \textsc{DecompIR}\footnote{The code is publicly available \url{https://sites.google.com/site/decompir/}.} \citep{Mullaney11}. 
The choice of \textsc{DecompIR} is driven by the fact that the code is popular to derive the IR properties of AGN host galaxies and is publicly available.
The aim here is not to compare the results obtained from CIGALE and \textsc{DecompIR} as they are two very different methods. 
Indeed, results from CIGALE are based on energy balance SED fitting using multi wavelength data from UV to submm, whereas \textsc{DecompIR} is based only on the IR. 
\textsc{DecompIR} uses a set of observed templates for the host and the AGN. 
The host galaxy templates are derived by using the \cite{Brandl06} SB sample, as well as four galaxies taken in the Revised Bright Galaxy Sample. 
These host templates are grouped into five different averages, spanning a wide range of possible observed SB IR SEDs. 
The AGN template is defined by effectively subtracting these host templates from the IR SEDs of local AGNs \citep{Mullaney11}.

To determine how successfully \textsc{DecompIR} separates the AGN component from the host emission, and thus estimates the SFR, we perform SED fitting with the five different SB templates and an AGN component using MIPS 24\microns\ to SPIRE 500\microns\ data, corresponding to 12 to 250\microns\ rest frame. 
For each mock galaxy, we also fit the IR SEDs, using only the five different SB components, i.e. without allowing any AGN contribution. 
This last test is performed to examine the value of adding an AGN component to the fit. 
 As a result, each mock galaxy  is tested with ten different configurations (i.e. 5 double components and 5 single component) for which $\chi^2$ are derived.

We then use an Akaike Information Criterion (AIC) to select the best model, regardless the number of components used to fit the data. 
The AIC is comparable to the more popular Bayesian Information Criterion (BIC). 
Although both criteria are derived in the same way, the main difference comes from the prior used which is inversely proportional to the number of models for the BIC, and a decreasing function of the number of models for the AIC.
However, some advanced studies comparing both criteria, showed that the AIC is more accurate in model selection than the BIC (see \cite{BurnhamAnderson04} and \cite{Yang05} for detailed comparisons).
For each mock galaxy, we submit the ten different $\chi^2$ to the AIC. 
We find that for the Type 1, intermediate type, and Type 2 AGNs, the AGN components improves the fit for 36, 21 and 47\% of the galaxies, respectively. 
For all of the other galaxies, whatever the type, a SB template is enough. 
We then estimate the AGN fraction, the $L_{IR}$, and the uncontaminated SFR, using \cite{Kennicutt98}.

Figure~\ref{decompir} shows the results, i.e. the variation of the estimation of the AGN contribution, the SFR and the $L_{IR}$ with the AGN contribution using \textsc{DecompIR}. 
We present the estimate of the $L_{IR}$ as the SFR is determined from the simple conversion of $L_{IR}$ using \cite{Kennicutt98}.
The AGN contribution is always underestimated (except for one point corresponding to $frac_{AGN}$ $\sim$20\% in the Type 2 sample). 
Confirming what observed with the CIGALE results, low fractions of AGN are difficult to constrain. 
The Type 1 AGN contribution is underestimated by at least 25\%, whereas intermediate type $frac_{AGN}$ is closer to the true value with an offset comprised between 15 and 50\%. 
The variation of the estimation of the AGN contribution of the Type 2 sample is however peculiar as it is not monotonic. 
Like the results obtained with CIGALE, the SFR is slightly underestimated but well recovered whatever the fraction of AGN for the Type 1 and intermediate type sample, and up to 70\% for the Type 2 sample. 
This is linked to the good recovery of the $L_{IR}$ with a small overestimation of a few percent in the absence of AGN, that does not evolve with $frac_{AGN}$ for the intermediate type sample, slightly increases up to $\sim$15\% for the Type 1 sample, and follow the peculiar behavior observed for the estimation of $frac_{AGN}$ for the Type 2 sample. 
However, despite problems in recovering the right contribution of the AGN, SFR of Type 1 and intermediate type AGNs are well recovered.

The estimate of the SFR and $L_{IR}$ of our Type 2 AGN sample by \textsc{DecompIR} is problematic for fractions larger than 40\%.
One explanation is that Type 2 templates from \cite{Fritz06} predict IR SED cooler than what was empirically obtained by \cite{Mullaney11}.
Thus, when \textsc{DecompIR} tries to fit the Type 2 catalogue created from \cite{Fritz06} models, the models attribute the excess of FIR emission produced by the Type 2 AGN from \cite{Fritz06} to star formation, leading to the offset on the $L_{IR}$ that we see in Figure~\ref{decompir}.
In order to test this point, we run \textsc{DecompIR} on the mock catalogue obtained with the Type 2 AGN template of \cite{Lusso13} used in Section~\ref{Lusso}.
The results are presented in purple in Figure~\ref{decompir}.
The SFR and $L_{IR}$ are no longer overestimated.
\textsc{DecompIR} provides a good estimate of the SFR with a small underestimation between 10 and 20\%, as well as for the $L_{IR}$ with an underestimation up to 20\% for the highest fraction.
Interestingly, the estimate on the contribution of the AGN to the $L_{IR}$ is well constrained and is similar to what obtained with CIGALE, i.e an overestimation of low fractions of about 50\%.

%=================================================================================
\section{Conclusions}

We simulate realistic SEDs of Type 1, intermediate type, and Type 2 AGNs hosts in order to evaluate the impact of the AGN emission on the host SED, and estimate the ability of SED fitting code to retrieve the physical properties of the galaxy.
CIGALE is used to model the SEDs from SFHs provided by the SAM \galform\ code.
Three samples are built, one with the emission of a Type 1AGN, one with the emission of an intermediate type AGN, and one with the emission of a Type 2 AGN, varying their contribution through the $frac_{AGN}$ parameter.
We use the SED fitting function of the recently updated CIGALE model to derive the stellar mass, star formation rate, and contribution of the AGN of each mock galaxy, assuming three popular shapes of SFH: 1$\tau$-dec, 2$\tau$-dec, and a delayed SFH.

In the absence of an AGN contribution, i.e. in normal galaxies, the SED fitting of our mock samples shows that all the three SFHs considered in this work provide good estimations of the stellar mass within $\sim$10\%. 
The best estimates are obtained using the 2$\tau$-dec model but at the expense of realistic ages of the stellar population, whereas the delayed model provide both.
Star formation rates are well recovered within 12\%.

For AGNs, the SED fitting of our mock samples shows that:
\begin{itemize}
	\item Stellar masses are overall well recovered with systematics up to 40\% (0.17\,dex) or better, depending on the $frac_{AGN}$ and spectral shape of the AGN component. This results is rather insensitive to photometric bands available as long as UV--MIR data are available. $M_*$ is therefore the most robust parameter that one can constraint for AGN hosts galaxies via broad-band photometric decomposition.
	\item The SFR suffers from systematic uncertainties up to 40-50\%  or better as long as FIR/submm data are available. Data-sets that are limited to the MIR cannot be used to constrain the SFR of AGN hosts.
	\item AGN/galaxy decomposition based on broad-band photometry can lead to significant overestimation of the AGN fraction and hence the inferred AGN luminosity in the case of weak AGN. 
\end{itemize}

We note the need for UV rest frame data to constrain the stellar mass and star formation rate of Type 1 and intermediate type AGNs.

The AGN emission has an influence on the slope of the MS depending on the SFH assumptions used for the SED fitting and the type of AGN .
As the AGN contribution can slightly bias the estimates of $M_*$ and SFR, the variety of AGN types and AGN intensities increases the dispersion of the MS.
Furthermore, the SFR-$M_*$ relation is more sensitive to the estimation of the stellar mass than of the SFR.

Finally, we use our mock samples to test a popular method used to disentangle the AGN emission from the host emission in the IR, \textsc{DecompIR} \citep{Mullaney11}, and find that, using the mock catalogues built from \cite{Fritz06} AGN templates, $frac_{AGN}$ is always underestimated but the SFR is well recovered for Type 1 and intermediate types of AGN, and overestimated in Type 2 AGNs when $frac_{AGN}>$35\%.
The overestimation of the SFR of Type 2 AGNs is due to the FIR prediction of Type 2 AGN from \cite{Fritz06} which is colder than what observed by \cite{Mullaney11}.
When using the Type 2 mock catalogue built from \cite{Lusso13}, \textsc{DecompIR} recovers well the SFR and $L_{IR}$ with an underestimation up to 20\%.

In this work, we have validated the use of broad-band SED fitting methods to derive the stellar mass and SFR of AGN host galaxies, as well as the contribution of this AGN to the host galaxy SED.
This analysis provides the foundation for future AGN multi-wavelength studies using CIGALE.
Indeed, in the near future, the eROSITA X-ray telescope will provide observations of about three millions of X-ray detected AGNs as well as samples of tens of thousands obscured AGNs.
These data will provide accurate studies of the relationship between accretion on the SMBH and the host galaxy properties. 
Multi-wavelength analysis using SED fitting will thus be needed to derive these properties.

%=================================================================================
\begin{acknowledgements}
We thank the referee for his/her detailed comments that helped improving the paper.
L.~C. warmly thanks M.~Boquien, Y.~Roehlly and D.~Burgarella for developing the new version of CIGALE on which the present work rely on.
L.~C. also thanks James Mullaney for useful discussions and suggestions, and Elisabeta Lusso for providing her SED templates.
This work benefited  from the {\sc  thales} project 383549 that  is jointly funded by the European Union and the Greek Government in the framework of the program ``Education and lifelong learning''.
\end{acknowledgements}
%%=================================================================================
\bibliographystyle{aa}
\bibliography{pcigale_galform_v5}
%=================================================================================
\appendix
\section{\label{iffitnoagn}The need for an AGN component to perform the SED fitting}

In this work, we use a SED fitting procedure allowing the possibility of using an AGN component.
To understand how omitting to use this component would affect the estimate of the stellar mass and SFR of the host galaxies, we perform a SED fitting, using the 2$\tau$-dec SFH model, and not allowing for the use of AGN templates.

\begin{figure*}%[!h] 
	\includegraphics[width=\textwidth]{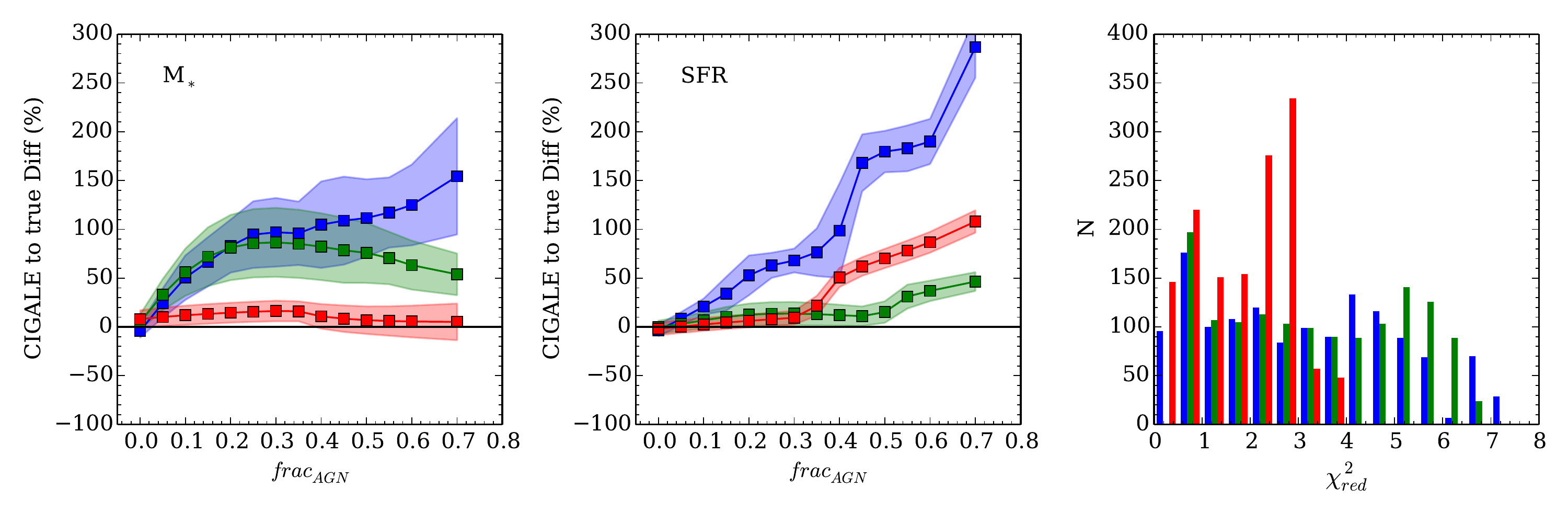}
 	\caption{ \label{fitnoAGN} Evolution of the CIGALE to true relative difference of $M_*$ and SFR with the fraction of AGN in the case where the AGN component is not taken into account in the fitting procedure. Left panel shows the results for the stellar mass, middle panel for the SFR, and the right panel present the $\chi^2_{red}$ distribution. Blue, green, and red data are for Type 1, intermediate type, and Type 2 AGNs, respectively.}
\end{figure*}

Figure~\ref{fitnoAGN} presents the fractional difference between the output parameters and the true ones as a function of the input $frac_{AGN}$.
The estimate of the stellar mass of the Type 2 sample is not perturbed by the AGN, as we discuss in this work. 
The NIR bands are not affected by the AGN emission, thus the stellar mass is well recovered, with or without the use of an AGN component.
However, Type 1 and intermediate type SEDs are rapidly affected by the AGN emission, and $M_*$ is overestimated by 50\% at $frac_{AGN}$=10\%.
This overestimation reaches up to 150\% for $frac_{AGN}$=70\% in the case of Type 1 AGNs.

The SFR in intermediate type and Type 2 AGNs is overestimated by 20\% up to $frac_{AGN}$=30-40\%.
This overestimation increases for higher fractions, especially in Type 2 AGNs where it can reach up to 100\%
The case of Type 1 AGNs is more critical, the SFR is rapidly strongly overestimated up to 300\% at $frac_{AGN}$=70\%.

The $\chi^2_{red}$ distributions shows higher value of the $\chi^2_{red}$ by a factor of $\sim$10 compared to the distribution from the fitting using the AGN component (Figure~\ref{sfhs}).

These results show that not taking into account an AGN component when performing broad band SED fitting of AGN host galaxies results in strong biases in the determination of physical properties such the stellar mass and the star formation rate.

\end{document}